\documentclass[journal]{IEEEtran}
\usepackage{amssymb}
\usepackage[cmex10]{amsmath}
\usepackage{stfloats}
\usepackage{graphicx}
\usepackage{subfigure}
\usepackage{tabularx}
\usepackage{epsfig,epsf,color,balance,cite}
\usepackage{verbatim}
\usepackage{url}
\usepackage{bm}

\newtheorem{theorem}{Theorem}

\newtheorem{lemma}{Lemma}

\usepackage{algorithm}
\usepackage{algorithmic}
\usepackage{color}
\definecolor{myc1}{rgb}{0,0,0}
\definecolor{myc2}{rgb}{0,0,0}
\begin{document}

\title{Optimal Fairness-Aware Time and Power Allocation in  Wireless Powered Communication Networks}
\author{
\IEEEauthorblockN{Zhaohui Yang, 
                  Wei Xu, \IEEEmembership{Senior Member, IEEE},
                  Yijin Pan, 
                  Cunhua Pan, \IEEEmembership{Member, IEEE} \\ and
                  Ming Chen, \IEEEmembership{Member, IEEE}
                  }
\thanks{This work was supported in part by the National Nature Science Foundation of China under grants 61372106, 61471114 \& 61221002, in part by the UK Engineering and Physical Sciences Research Council under Grant EP/N029666/1, in part by the Six Talent Peaks project in Jiangsu Province under grant GDZB-005, and in part by the Scientific Research Foundation of Graduate School of Southeast University under Grant YBJJ1650. (\emph{Corresponding authors: Wei Xu; Cunhua Pan }.)}
\thanks{Z. Yang, W. Xu, Y. Pan and M. Chen are with the National Mobile Communications Research Laboratory, Southeast University, Nanjing 210096, China  (Email: yangzhaohui@seu.edu.cn, wxu@seu.edu.cn, panyijin@seu.edu.cn, chenming@seu.edu.cn).}
 \thanks{C. Pan is with the School of Electronic Engineering and Computer Science, Queen Mary, University of London, London E1 4NS, U.K. (Email: c.pan@qmul.ac.uk).}
}
\maketitle

\begin{abstract}
In this paper, we consider the sum $\alpha$-fair utility maximization problem for joint downlink (DL) and uplink (UL) transmissions of a wireless powered communication network (WPCN) via time and power allocation.
In the DL, the users with energy harvesting receiver architecture decode information and harvest energy based on simultaneous wireless information and power transfer.
While in the UL, the users utilize the harvested energy for information transmission, and {\color{myc1}{harvest energy when other users transmit UL information}}.
We show that the general sum $\alpha$-fair utility maximization problem can be transformed into an equivalent convex one.
Tradeoffs between sum rate and user fairness can be balanced via adjusting the value of $\alpha$.
In particular, for zero fairness, i.e., $\alpha=0$, the optimal allocated time for both DL and UL is proportional to the overall available transmission power.
Tradeoffs between sum rate and user fairness are presented through simulations.
\end{abstract}


\IEEEpeerreviewmaketitle

\section{Introduction}

Energy harvesting has been recognized as a key technology for wireless communications \cite{7317504,7081084}.
By using energy harvesting, users are able to harvest wireless energy from environment or radio frequency (RF) signals \cite{6284703,5513714,4595260,6623062,Zhang2016EE,Zhang2016Art}.
Due to this distinction, energy harvesting technology can prolong the lifetime of energy-constrained networks, such as sensor, ad-hoc and Internet of Things (IoT) \cite{7448818,7120024,5522465,7801070}.

Since energy harvesting from the environment is an uncontrollable process,
wireless powered communication networks (WPCNs), where users are powered
over the air by dedicated wireless power transmitters, have attracted great interest in both academia and industry \cite{7462480,6957150,7744829,6678102,7037291,7374746,7328715,7782791}.
In \cite{6678102}, the optimal time allocation scheme was investigated for WPCNs where multiple users harvest energy from base station (BS) downlink (DL) broadcasting and utilize the harvested energy to transmit information in the uplink (UL).
The time allocation problem for WPCNs was studied in \cite{7037291} , where the BS equipped with two antennas can simultaneously broadcast energy wirelessly in the DL and receive information in the UL.
The optimal time allocation was efficiently identified by a simple line search method in \cite{7374746} for a multiuser multiple-input multiple-output (MIMO) WPCN.
To further maximize the sum rate in WPCNs, \cite{7328715} jointly optimized BS broadcasting power and the time sharing among users.
Moreover, the energy efficiency maximization problem for WPCNs via joint time and power allocation was investigated in \cite{7782791}.
According to \cite[Section~ III-B]{6750095} and \cite[Section~III-C]{6472116},
data transmission in the DL is considered to provide information exchange in IoT, which is an important application to WPCNs.
The above existing works \cite{6678102,7037291,7374746,7328715,7782791} all assumed that users only harvest energy in the DL even though the wireless information transmission in the DL is required in most practical applications.

{\color{myc1}{
A centralized wireless time-division duplexing  orthogonal frequency division multiple access network with energy harvesting was considered in \cite{6884177}, where different users occupy different subchannels in the DL and UL.
However, the DL and UL time fractions were set as fixed in \cite{6884177}, even though the DL and UL time fractions can be optimized to further improve the system performance.
In \cite{6977947}, a DL time-slotted simultaneous wireless information and power transfer network was considered, where a single user is scheduled for information transmission in each slot and the remaining users only opportunistically harvest energy from received signals.
Considering both DL and UL, \cite{7848950} investigated a simultaneous wireless information and power transfer network with only one DL user and multiple UL users.
Since channel gains were assumed to be the same in different blocks in \cite{7848950}, equal power allocation is proved to be optimal, which simplifies the original link and power allocation problem to a merely link allocation problem.
}}

The sum rate and the fairness of all users are two crucial performance metrics.
Both are similar to the two ends of a seesaw in that improving one will often decrease the other \cite{6816520}.
Toward the two ends, the sum rate and max-min rate problems for WPCNs were studied in \cite{6678102}.
Since maximizing sum rate usually results in the most unfair case and the max-min optimization often leads to the lowest sum rate \cite{7111370},
an efficient tradeoff between sum rate and fairness is needed.
Recently, \cite{7875145} and \cite{7867035} investigated proportional fairness problems for WPCNs via time allocation.
However, proportional fairness can only be regarded as a special tradeoff point.
To offer a smooth tradeoff between sum rate and fairness, sum $\alpha$-fair utility maximization problem \cite{6816520} can be a powerful candidate.
To the best of our knowledge, the sum $\alpha$-fair utility maximization problem for WPCNs is rarely considered.

In this paper, we consider joint time and power allocation for WPCNs to maximize sum $\alpha$-fair utility.
Different from prior works \cite{6678102,7037291,7374746,7328715,7782791}, which focused on wireless energy transfer in the DL and information transmission in the UL,
in this paper users simultaneously receive wireless information and harvest energy in both DL and UL.
{\color{myc1}{
Different from \cite{7848950}, we consider a general WPCN serving multiple users with quasi-static flat-fading channels in both DL and UL.}}

The main contributions of this paper are summarized as follows:
\begin{itemize}
  \item  We formulate the sum $\alpha$-fair utility maximization problem for WPCNs.
      The general sum $\alpha$-fair utility maximization problem is proved to be a convex problem.
      The maximal sum $\alpha$-fair utility can be achieved by occupying all available time. 
  \item  For zero fairness with given DL transmission power and split power, i.e., $\alpha=0$, the optimal solution is obtained in closed form.
      Since power loss due to path loss can be minimized, it is optimal for at most one user in each epoch to transmit with positive power in the UL.
      Besides, the optimal allocated time is proportional to the overall power available for transmission.
  \item  When $\alpha>0$, we propose iterative time and power allocation algorithms with low complexities, each step of which obtains the optimal solution in closed form.
      For common fairness and max-min fairness, the optimal split power for each user in the DL increases with its allocated time, and the optimal transmission power for each user in the UL also increases with the  allocated time.
\end{itemize}

This paper is organized as follows.
In Section $\text{\uppercase\expandafter{\romannumeral2}}$, we introduce the system model and provide the formulation of sum $\alpha$-fair utility maximization problem.
Section $\text{\uppercase\expandafter{\romannumeral3}}$ provides
the optimality of full time and maximal power, and proposes optimal time and power allocation schemes.
Some numerical results are displayed in Section $\text{\uppercase\expandafter{\romannumeral4}}$
and conclusions are finally drawn in Section $\text{\uppercase\expandafter{\romannumeral5}}$.

\section{System Model and Problem Formulation}
\label{section2sys}
\subsection{System Model}
\begin{figure}
  \centering
   \includegraphics[width=3.5in]{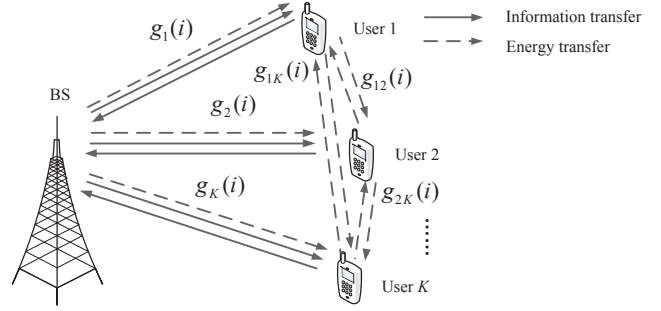}\\
    \caption{System model. \label{fig1}}
\end{figure}

Consider a WPCN with simultaneous wireless information and power transfer in both DL and UL, as shown in Fig.~\ref{fig1}.
In this network, there exist one BS and $K$ users denoted by the set $\mathcal K=\{1, \cdots, K\}$.
The network utilizes time division multiple access (TDMA), and the time is divided into $M$ TDMA frames (also referred to as the epochs) of equal duration $T$.
It is assumed that the BS and all users are equipped with one single antenna each.
Channel reciprocity \cite{Qiu2007Channel} holds for the DL and UL.
It is further assumed that all channels are quasi-static flat-fading,
i.e., the channel is constant during a single epoch but changes independently from one epoch to the next.
The channel gain between the BS and user $k$ in epoch $i$ is denoted by $g_k(i)$, $\forall k \in \mathcal K$, $i\in\mathcal M=\{1, \cdots, M\}$.
{\color{myc1}{Assume that full channel state information (CSI) is available at the BS. This can be realized by setting up CSI feedback channels from all users with predetermined codebooks at both sides.}}

The network adopts a \emph{harvest-then-transmit} protocol as shown in Fig.~2.
{\color{myc1}{During the DL phase in epoch $i$, a fraction of time $m_k(i) T$, $0\leq m_k(i)<1$, is assigned to user $k$ for information transmission, $\forall k \in \mathcal K$.
Each user $k$ harvests energy when the BS transmits information to other users, i.e., user $k$ harvests energy within the time fraction $\sum_{l\in\mathcal K\setminus \{k\}} m_l(i) T$.
The fraction of time allocated to user $k$ during the UL phase in epoch $i$ is denoted by $n_k(i) T$, $0\leq n_k(i) <1$, $\forall k \in \mathcal K$.
Each user also harvests energy in the UL when other users transmit UL information.
Since $m_k(i)$ and $n_k(i)$ represent the time portions allocated to user $k$ for information transformation in the DL and UL, respectively, we have
\begin{equation}\label{sys2eq1}
\sum_{k\in \mathcal K} (m_k(i)+n_k(i))T \leq T, \quad \forall i \in \mathcal M.
\end{equation}}}
For convenience, we assume a normalized unit block time $T=1$ in the following.
Without causing ambiguity, we can use both the terms of energy and power interchangeably.

\begin{figure*}
  \centering
  \includegraphics[width=4.45in]{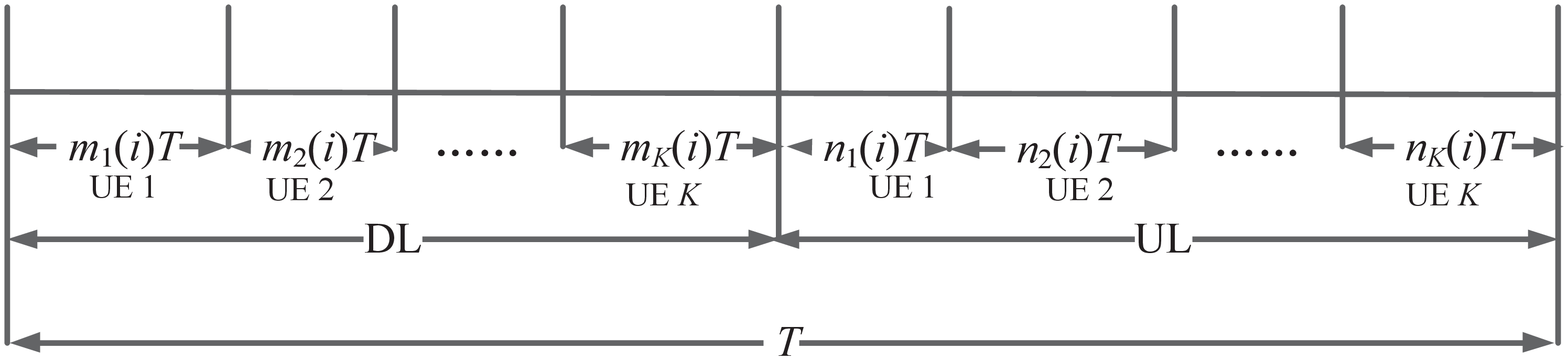}\\
  \caption{The \emph{harvest-then-transmit} protocol.}
\end{figure*}

During the DL phase in epoch $i$, the transmission power of the BS for user $k$ is denoted by $p_k(i)$.
The received noise at each user is modeled as a circularly symmetric complex Gaussian (CSCG) random variable with zero mean and variance $\sigma^2$.
{\color{myc1}{The received signal at user $k$ is processed by a power splitter, where a ratio $1-\rho_k(i)$ of power is split to its energy receiver and the remaining $\rho_k(i)$ part is split to its information receiver, with $0 \leq \rho_k (i) \leq 1$.}}
The achievable DL rate of user $k$ in bits/s/Hz for the information receiver can be expressed as
\begin{equation}\label{sys2eq3}
r_k^{\text{DL}}(i)= m_k(i) \log_2 \left(
1+\frac{g_k(i) \rho_k(i) p_k(i)}
{\Gamma \sigma^2}
\right), \quad \forall k \in \mathcal K,
\end{equation}
where $\Gamma$, namely system margin, is the signal-to-noise gap from the additive white Gaussian noise due to practical modulation and coding implementation.

{\color{myc1}{In epoch $i$, the achievable UL rate of user $k$ in bits/s/Hz is given by
\begin{equation}\label{sys2eq6}
r_k^{\text{UL}}(i)=
n_k (i) \log_2 \left(1+
\frac{
g_k (i) \bar p_k (i) }
{\Gamma \sigma^2}\right)
, \quad \forall k \in \mathcal K,
\end{equation}
where $\bar p_k(i)$ is the UL transmission power of user $k$.}}

It is assumed that the energy harvested due to the received noise is negligible.
{\color{myc2}{
The energy harvested by power receiver of user $k$ between user $k$'s UL information transmission in epoch $i-1$ and user $k$'s UL information transmission in epoch $i$ is given by
\begin{eqnarray}\label{sys2eq4}
  && \!\!\!\!\!\!\!\!\!\!\!\!E_k(i) \! =\!
  \underbrace{\zeta g_k(i) \!\sum_{l\in\mathcal K \setminus \{k\}}\! m_l (i) p_l (i)  \!+ \! \zeta g_k(i) m_k(i) (1\!-\!\rho_k (i)) p_k(i) } _{\text { harvested energy in the DL } }
  \nonumber \\
  && \!\!\!\!\!\!\!\!\!\!\!\!
  \!+\!
  \underbrace { \zeta_0 \!\sum_{l=1} ^{k-1}\! g_{lk} (i) n_l(i) \bar p_l (i)\! +\!\! \zeta_0\!\! \sum_{l=k+1}^K \!g_{lk} (i-1) n_l (i-1) \bar p_l (i-1)}_ {\text{ harvested energy in the UL  }} 
\end{eqnarray}
for all $ k \in \mathcal K$,
where $g_{lk}(i)$ is the channel gain between user $l$ and user $k$ in epoch $i$, $\bar p_k(0)=0$ for all $k\in \mathcal K$,
$\zeta \in(0,1]$ is the conversion efficiency of the energy harvested from the BS,
and $\zeta_0 \in[0,1]$ represents the conversion efficiency of the energy harvested from the other users.
Note that the case $\zeta_0=0$ represents that the user cannot harvest the energy from the other users.}}

\subsection{Problem Formulation}

Define the $\alpha$-fair utility function \cite{879343} by $U_{\alpha}(x)$, which is given as
\begin{equation}\label{sys2eq7}
\begin{aligned}
&U_{\alpha}(x)
=
\begin{cases}
\ln(x)& \mbox{if $\alpha$=1}\\
\frac{1}{1-\alpha} x^{1-\alpha}& \mbox{if $\alpha\geq 0$, $\alpha\neq1$}.
\end{cases}
\end{aligned}
\end{equation}
In particular, the utility function $U_{\alpha}(x)$ represents zero fairness with $\alpha=0$, proportional fairness with $\alpha=1$, harmonic mean fairness with $\alpha=2$, and max-min fairness with $\alpha=+\infty$.

We aim at network utility maximization via time and power allocation.
The sum $\alpha$-fair utility maximization problem can be formulated as
\begin{subequations}\label{sys2max1}
\begin{align}
\mathop{\max}_{\pmb{m}, \pmb n, \pmb{p}, \bar {\pmb p}, \pmb \rho \geq \pmb 0}\quad
& \sum_{k\in \mathcal K}    U_{\alpha} \left(\frac{1} M \sum_{i\in\mathcal M}r_k^{\text{DL}}(i)\right)
\nonumber\\
& +  \sum_{k\in \mathcal K}  U_{\alpha} \left( \frac 1 M \sum_{i\in\mathcal M} r_k^{\text{UL}} (i) \right)
 \\
\textrm{s.t.}\qquad  \:\:
&
\sum_{k\in \mathcal K} m_k(i)+n_k(i) \leq 1, \quad \forall i \in \mathcal M \\
&{\color{myc2}{ p_k(i) \leq P_{\max}, \quad \forall i \in \mathcal M, k \in \mathcal K}}\\
&\frac{1} M \sum_{i\in\mathcal M } \sum_{k \in \mathcal K} m_k(i) p_k(i) \leq P_{\text {avg  } } \\
& \sum_ {j=1}^i n_k (i) \bar p_k (i) \leq \sum_ {j=1}^i E_k(i), \quad \forall  i \in \mathcal M, k \in \mathcal K \\
& \pmb \rho \leq \pmb 1,
\end{align}
\end{subequations}
where $\pmb m=\{m_k (i)\}_{i\in\mathcal M, k \in \mathcal K}$, $\pmb n=\{n_k (i)\}_{i\in\mathcal M, k \in \mathcal K}$, $\pmb p=\{p_k (i)\}_{i\in\mathcal M, k \in \mathcal K}$, $\bar{ \pmb p}= \{\bar p_k (i)\}_{i\in\mathcal M, k \in \mathcal K}$, $\pmb \rho=\{\rho_k (i)\}_{i\in\mathcal M, k \in \mathcal K}$,
{\color{myc2}{$P_{\max}$ is the maximal instantaneous transmission power of the BS}}, and
$P_{\text {avg} }$ is the maximal average transmission power of the BS.

\section{Optimal Time and Power Allocation}
It is easily verified that problem (\ref{sys2max1}) is nonconvex due to the nonconvex constraints in (\ref{sys2max1}c)-(\ref{sys2max1}e).
In order to make the problem more tractable,
we first reformulate problem (\ref{sys2max1}) by introducing a set of new nonnegative variables:
$q_k(i) = m_k (i) p_k (i)$, $\bar q_k (i) =n_k (i) \bar p_k (i)$, $v_k (i) = m_{k} (i) \rho_k (i) p _k (i)$, $\forall i \in \mathcal M, k \in \mathcal K$.
Based on the above definitions, $q_k(i)$ and $\bar q_k (i)$ can be respectively viewed as DL and UL transmission powers, while $v_k(i)$ is viewed as split power.
Then, problem (\ref{sys2max1}) is equivalent to
\begin{subequations}\label{eem3max0}
\begin{align}
\mathop{\max}_{ \pmb{m}, \pmb n,\pmb{q} , \bar { \pmb q }, \pmb v \geq \pmb 0 } \quad \!\!\!
&
 \sum_{k\in \mathcal K}    U_{\alpha} \left(\frac{1} M \sum_{i\in\mathcal M} \bar r_k^{\text{DL}}(i)\right)
 \nonumber \\ & +  \sum_{k\in \mathcal K}  U_{\alpha} \left( \frac 1 M \sum_{i\in\mathcal M} \bar r_k^{\text{UL}} (i) \right)
\\
\textrm{s.t.}\qquad  \:\:
&
\sum_{k\in \mathcal K} m_k(i)+n_k(i) \leq 1, \quad \forall i \in \mathcal M \\
&{\color{myc2}{q_k(i) \leq P_{\max} m_k(i)}}, \quad \forall i \in \mathcal M, k \in \mathcal K\\
&\frac{1} M \sum_{i\in\mathcal M } \sum_{k \in \mathcal K} q_k(i) \leq P_{\text {avg  } } \\
& \sum_{j=1}^i   \bar q_k (i) \leq \sum_ {j=1}^i \bar E_k(i), \quad \forall  i \in \mathcal M, k \in \mathcal K \\
& v_k (i) \leq q_k (i) , \quad \forall  i \in \mathcal M, k \in \mathcal K,
\end{align}
\end{subequations}
where $\pmb q=\{q_k (i)\}_{i\in\mathcal M, k \in \mathcal K}$, $\bar{ \pmb q}= \{\bar q_k (i)\}_{i\in\mathcal M, k \in \mathcal K}$, $\pmb v=\{v_k (i)\}_{i\in\mathcal M, k \in \mathcal K}$,
\begin{equation}\label{eem3max0eq1}
\bar r_k^{\text{DL}} (i)=
m_k (i) \log_2 \left(1+\frac{g_k(i) v_k(i) }{\Gamma \sigma^2 m_k(i)} \right),
\end{equation}
\begin{equation}\label{eem3max0eq1_2}
\bar r_k^{\text{UL}} (i) =
n_k (i) \log_2\left(1+
\frac{
 g_k (i) \bar q_k (i)}
{\Gamma \sigma^2 n_k (i) }\right),
\end{equation}
and
\begin{eqnarray}\label{eem3max0eq3}
\bar E_k(i)=&& \!\!\!\!\!\!\!\!
  \zeta g_k(i) \sum_{l\in\mathcal K } q_l (i)  -  \zeta g_k (i) v_k(i)
+
   {\color{myc2}{\zeta_0}} \sum_{l=1} ^{k-1} g_{lk} (i) \bar q_l (i)
  \nonumber \\     && \!\!\!\!\!\!\!\!
 + {\color{myc2}{\zeta_0}} \sum_{l=k+1}^K g_{lk} (i-1)  \bar q_l (i-1).
\end{eqnarray}

\begin{theorem}
The objective function of problem (\ref{eem3max0}) is concave with respect to (w.r.t.) $(\pmb m, \pmb n, \bar {\pmb q}, \pmb v)$.
\end{theorem}

\itshape \textbf{Proof:}  \upshape
Please refer to Appendix A.
 \hfill $\Box$

To solve problem (\ref{eem3max0}), we have the following lemma.
\begin{lemma}
The optimal ($\pmb m^*, \pmb n^*$) of problem (\ref{eem3max0}) satisfies $\sum_{k\in \mathcal K}(m_{k}^*(i)+n_k^*(i))= 1$, $\forall i \in\mathcal M$.
\end{lemma}

\itshape \textbf{Proof:}  \upshape
Please refer to Appendix B.
 \hfill $\Box$

According to Lemma 1, transmitting with full time is optimal.
This is also intuitively reasonable because both DL and UL rates of users can always be enhanced with the increase of time.
From Theorem 1, the objective function of problem (\ref{eem3max0}) is concave.
Since the constraints of (\ref{eem3max0}) are all linear, problem (\ref{eem3max0}) is a convex problem, which can be solved by the popular interior point method (IPM) \cite{boyd2004convex}.
For finite $M$, solving (\ref{eem3max0}) with constraints (\ref{eem3max0}e) is difficult and requires non-causal CSI knowledge.
According to \cite{Zlatanov2017Aympto} and \cite{Hadzi2017Opportunistic}, one user with unlimited storage capacity can transmit with the desired transmit power in almost all epochs, if the number of epochs $M$ satisfies $M \rightarrow \infty$.
As a result, constraints (\ref{eem3max0}e) can be simplified when $M \rightarrow \infty $, as the following constraints:
\begin{equation}\label{eem3eq1}
\sum_ {i=1} ^M \bar q_k (i) \leq \sum_ {i=1} ^M \bar E_k(i), \quad \forall k \in \mathcal K.
\end{equation}
Note that the left hand side of (\ref{eem3eq1}) is the total energy consumed for transmission of user $k$, while the right hand side is the total energy harvested by user $k$.

In the following, we therefore replace (\ref{eem3max0}e) with simplified constraints (\ref{eem3eq1}). We consider three cases separately: zero fairness with $\alpha=0$, common fairness with $0<\alpha<+\infty$, and max-min fairness with $\alpha=+\infty$.
{\color{myc2}{
Having obtained the optimal solution ($\pmb m^*, \pmb n^*, \pmb q^*, \bar {\pmb q}^*, \pmb v^*$) to the equivalent convex problem (\ref{eem3max0}), the optimal solution ($\pmb{m}, \pmb n, \pmb{p}, \bar {\pmb p}, \pmb \rho$) to the original problem  (\ref{sys2max1}) is thus given by
\begin{equation}
m_k(i) = m_k^* (i), n_k (i) =n_k^* (i), \quad \forall i \in \mathcal M, k \in \mathcal K,
\end{equation}
and
\begin{equation}
p_k (i)=\frac{q_k^*(i)}{m_k^*(i)},
\bar p_k(i)=\frac{\bar q_k^*(i)}{n_k^*(i)},
\rho_k (i) =\frac{\bar v_k^*(i)}{q_k^*(i)}
\end{equation}}}
for all $ i \in \mathcal M, k \in \mathcal K$.
\subsection{Zero Fairness with $\alpha=0$}
When $\alpha=0$, problem (\ref{eem3max0}) becomes the zero fairness problem:
\begin{subequations}\label{eem3max1}
\begin{align}
\!\!\mathop{\max}_{ \pmb{m}, \pmb n,\pmb{q} , \bar {\pmb q}, \pmb v }\quad\!\!\!
&
{\frac{1} M \sum_{i\in\mathcal M}\sum_{k\in \mathcal K}   (\bar r_k^{\text{DL}}(i)  +  \bar r_k^{\text{UL}} (i) )}
\\
\textrm{s.t.}\quad  \:\!\!
&(\ref{eem3max0}b), (\ref{eem3max0}c), (\ref{eem3max0}d), (\ref{eem3eq1}), (\ref{eem3max0}f).
\end{align}
\end{subequations}

\begin{theorem}
With given ($\pmb q, \pmb v$), the optimal time and UL transmission power allocation to problem (\ref{eem3max1}), denoted by ($\pmb m^*, \pmb n^*,  \bar {\pmb q} ^* $), is given by
\begin{equation}\label{eem3eq2}
m_k^* (i)  =\left. \frac{g_k (i) v_k (i) } { \tilde C(i) } \right|_{\frac{q_k(i)}{P_{\max}}},
n_k^* (i) =\frac{ g_k (i) \bar q_k^* (i) } { C(i) }
\end{equation}
for all $ i \in \mathcal M, k \in \mathcal K$,
\begin{equation}\label{eem3eq3}
\bar q_{w (i) }^* (i) = \left\{
\begin{array}{ll}
\!\!  0 & \text{if} \; \tilde C(i) > D_{ w(i) } (i)   \\
\!\! \hat q_{w (i) }^* (i) & \text{if} \; \tilde C(i) \leq D_{ w(i) } (i)
\end{array}
\right.,\quad \forall i\in\mathcal M,
\end{equation}
and
\begin{equation}\label{eem3eq2_2}
\bar q_t^* (i) = 0, \quad \forall i\in\mathcal M, t \in \mathcal K \setminus \{ w(i) \},
\end{equation}
where  $ C(i) = \tilde C(i) $ if $\tilde C(i) > D_{ w(i) } (i) $,  $C(i) = D_{ w(i) } (i)$ if $\tilde C(i) \geq D_{ w(i) } (i) $,
$\tilde C(i)$ is the root of equation
\begin{equation}
\sum_{k\in \mathcal K}
\left. \frac{g_k (i) v_k (i) } { \tilde C(i) } \right|_{\frac{q_k(i)}{P_{\max}}} =1,
\end{equation}
$w (i) = \arg \min_ {k\in\mathcal K} D_ k (i) $, $D_{k} (i)$ is defined in (\ref{appentheorem1kkt5_7}),
$\hat q_{w (i) }^* (i) =\frac{ D_{ w(i) } (i) - \sum _{ k \in \mathcal K } g_k (i) v_k (i)|_{ \frac{q_k(i) D_{w(i)}} {P_{\max}}}} {g_{w(i)} (i)}$, and
$\mu$ and $\pmb \nu= \{\nu_k\}_{k \in \mathcal K}$ are the nonnegative Lagrange multipliers corresponding to (\ref{eem3max0}d) and (\ref{eem3eq1}), respectively.

With given ($\pmb m, \pmb n, \bar {\pmb q} $), the optimal DL transmission power and split power ($\pmb q^*, \pmb v^*$) is given by
{\color{myc2}{
\begin{equation}\label{eem3eq2_3}
q_k^* (i) = \left\{
\begin{array}{ll}
 \! \! v_k^* (i) & \text{if} \; \frac{\mu} { M } > \zeta \nu_k\sum _{l \in \mathcal K} g_l (i)  \\
\!\! P_{\max} m_k(i) & \text{if} \; \frac{\mu} { M } \leq  \zeta \nu_k\sum _{l \in \mathcal K} g_l (i)
\end{array}
\right.
\end{equation}
}}
and
\begin{equation}\label{eem3eq5}
v_k ^* (i) = \left. \left[ \frac{   m_k (i) }{ (\ln 2) \zeta \nu_k   g_{k } (i) } - \frac{  \Gamma \sigma^2 m_k (i) } { g_k (i) } \right] \right|^ {P_{\max} m_k(i)} _0
\end{equation}
for all $ i \in \mathcal M, k \in \mathcal K$.
\end{theorem}

\itshape \textbf{Proof:}  \upshape
Please refer to Appendix C.
 \hfill $\Box$

From Theorem 2, the allocated time in each epoch is proportional to the overall power available for both DL and UL transmissions.
It can be found that at most one user $w(i)$ in the UL desires positive power $\bar q_{w(i)}(i)$ to transmit information to the BS.
According to Theorem 2, problem (\ref{eem3max1}) can be effectively solved via iteratively optimizing ($\pmb m, \pmb n, \bar {\pmb q} $) with given ($\pmb q, \pmb v$) and updating ($ \pmb q ,\pmb v$) with fixed ($\pmb m, \pmb n, \bar {\pmb q} $), which fortunately has closed-form expression in each iteration.
{\color{myc1}{Since problem (\ref{eem3max1}) is convex based on Theorem 1, the iterative algorithm yields a locally optimal solution, which is also the globally optimal solution to convex problem (\ref{eem3max1}).
}}
Practical implementation of the zero fairness protocol at the BS is summarized in Algorithm 1.
In practice, the values of $ \mu $ and $ \{ \nu_{k} \} _{\forall k \in \mathcal K }  $ may not be available in advance, and we adopt the stochastic gradient descent method \cite{boyd2004ee364a} as in Algorithm 1, where $ \{ \lambda_ k\}_{ k\in \mathcal K \cup \{0\}} $ are some step sizes.

{\color{myc1}{
Note that the iterative steps (Step 4 and Step 5) in Algorithm 1 always converge due to the fact that (i) some variables are globally optimized with other variables fixed in each step, which shows that the objective function (\ref{eem3max1}a) is nondecreasing in each step; (ii) the objective function (\ref{eem3max1}a) is upper-bounded by
\begin{eqnarray}
&&\!\!\!\!\!\!\!\!\!\!\!\quad
{\frac{1} M \sum_{i\in\mathcal M}\sum_{k\in \mathcal K}   (\bar r_k^{\text{DL}}(i)  +  \bar r_k^{\text{UL}} (i) )}
 \nonumber\\&&\!\!\!\!\!\!\!\!\!\!\!
\overset{(\text a)}{=} \frac{1} M \sum_{i\in\mathcal M}\sum_{k\in \mathcal K}     \left( m_k (i) \log_2 \left(1+\frac{g_k (i) v_k (i)}{\Gamma \sigma^2 m_k (i)} \right)
\right.
 \nonumber\\&&\!\!\!\!\!\!\!\!\!\!\!
\left.
\quad +n_k (i) \log_2\left(1+
\frac{
\zeta g_k (i) \bar q_k (i) }
{\Gamma \sigma^2 n_k (i) }\right)
 \right)
\nonumber\\&&\!\!\!\!\!\!\!\!\!\!\!
\overset{(\text b)}{\leq}\frac{1} M \sum_{i\in\mathcal M}\sum_{k\in \mathcal K}     \left(   \log_2 \left(1+\frac{g_k (i) P_{\max}}{\Gamma \sigma^2  } \right)
\right.
  \nonumber\\&&\!\!\!\!\!\!\!\!\!\!\!
\left.
\quad
 +  \log_2\left(1+
\frac{
\zeta g_k (i) \bar P_{\max}  }
{\Gamma \sigma^2  }\right)
 \right),
\end{eqnarray}
where equality (a) follows from (\ref{eem3max0eq1}) and (\ref{eem3max0eq1_2}),
and inequality (b) holds because $0 \leq m_k(i), n_k(i) \leq 1$, $0\leq v_k (i) \leq q_k (i) \leq P_{\max}$,
$\bar P_{\max}$ determined by (\ref{eem3eq1}) is the maximal value of $\bar q_k (i)$,
and both $\bar r_k^{\text{DL}}(i)$ and $ \bar r_k^{\text{UL}} (i)$ increase with ($\pmb m,  {\pmb n}, \bar {\pmb q}, \pmb v$) according to Appendix B.}}

\begin{algorithm}[h]
\caption{: {\color{myc1}{Zero Fairness Protocol Implementation at the BS}} }
\begin{algorithmic}[1]
\STATE Initialize $\pmb v$, $ \mu $, and $\pmb \nu$ ; Set $ \{ \lambda_ k\}_{ k\in \mathcal K \cup \{0\}} $ and $i=1 $.
\REPEAT
\REPEAT
\STATE Obtain the optimal $\{m_{k} (i), n _k (i), \bar q_{k} (i) \}_{\forall k \in \mathcal K}$ with fixed $\{q _k (i), v_{k} (i)\}_{\forall k \in \mathcal K}$.
\STATE Obtain the optimal $\{q _k (i), v_{k} (i)\}_{\forall k \in \mathcal K}$ with fixed $\{m_{k} (i), n _k (i), \bar q_{k} (i) \}_{\forall k \in \mathcal K}$.
\UNTIL objective function (\ref{eem3max1}a) converges
\STATE Inform $\{ n _k (i), \bar q _k (i)/n_k(i), v_ k (i) /\bar q_k(i) \}_{\forall k \in \mathcal K}$ to all users, and broadcast RF energy at $p_k(i)=\frac{q_k (i)} { m _ k ( i ) }$ during $m_k (i)$, $\forall k \in \mathcal K$.
\STATE Update $\mu= \mu + \lambda_0  \left( \frac{1} i \sum_{t=1 } ^i \sum_{k \in \mathcal K} q_k(t) - P_{\text {avg  } } \right) $.
\STATE Update  $\nu_k= \nu_k +   \lambda_k    \left( \frac {1} i \sum_ {t=1} ^i \bar q_k (i) -  \frac {1} i  \sum_ {t=1} ^i \bar E_k(t)  \right) $, $\forall k \in \mathcal K$.
\STATE Set $i=i+1$.
\UNTIL $i \leq M$
\end{algorithmic}
\end{algorithm}

\subsection{Common Fairness with $0<\alpha<+\infty$}
Although the sum $\alpha$-fair utility maximization problem (\ref{eem3max0}) is convex, it is hard to obtain the optimal solution in closed form through directly solving the KKT conditions for an arbitrary $\alpha \in(0,+\infty)$.
In the following, we adopt an alternating method to solve problem (\ref{eem3max0}) by iteratively optimizing time sharing factor  with fixed   transmission power and split power, updating transmission power and split power with fixed time sharing factor.
In each step of the proposed alternating method, we can fortunately obtain closed-form optimal solutions to time sharing factor and power.

\begin{theorem}
Given ($\pmb q, \bar {\pmb q}, \pmb v$), the optimal time sharing factor  to problem (\ref{eem3max0}), denoted by ($\pmb m^*, \pmb n^*$), is determined by
\begin{equation}\label{Optalphamaxeq2}
m_k^* (i) \! =\! \left. - \frac {g_k (i) v_k (i)}{ \Gamma \sigma^2} \!\left(\!
\frac{1} { W(- \text e ^{-1 - (\ln 2) ( \bar R_k ^{\text {DL} } ) ^{\alpha } \phi (i) } ) }\! + \! 1  \!\right)\!
\right|_{\frac{q_k(i)}{P_{\max}}}
\end{equation}
and
\begin{equation}\label{Optalphamaxeq2_2}
n_k^* (i)  = \left. - \frac {g_k (i) \bar q_k (i)}{ \Gamma \sigma^2} \left(
\frac{1} { W(- \text e ^{-1 - (\ln 2) ( \bar R_k ^{\text {UL} } ) ^{\alpha } \phi (i) } ) } +1  \right)
\right|_{0}
\end{equation}
for all $i\in\mathcal M, k \in \mathcal K$,
where $W(\cdot)$ is Lambert-W function, 
$\bar R_k^{\text {DL}}$ and $\bar R_k ^{\text{UL}}$ are respectively the average rates over $M$ epochs in the DL and UL,
\begin{equation}\label{appenDeq1}
\bar  R_k ^{\text {DL}} = \frac{1} M \sum_{i\in\mathcal M} m_k (i) \log_2 \left(1+\frac{g_k(i) v_k(i) }{\Gamma \sigma^2 m_k(i)} \right),
\end{equation}
\begin{equation}\label{appenDeq1_2}
\bar   R_k ^{\text {UL}} =  \frac 1 M \sum_{i\in\mathcal M}n_k (i) \log_2\left(1+
\frac{
 g_k (i) \bar q_k (i)}
{\Gamma \sigma^2 n_k (i) }\right),
\end{equation}
and $\phi(i)$ is the root of the following equation
\begin{eqnarray}\label{Optalphamaxeq2_3}
&&\!\!\!\!\!\!\!\!\!\!\!\! \!\!  \sum_{ k \in \mathcal K }
\left. - \frac {g_k (i) v_k (i)}{ \Gamma \sigma^2} \left(
\frac{1} { W(- \text e ^{-1 - (\ln 2) ( \bar R_k ^{\text {DL} } ) ^{\alpha } \phi (i) } ) } +1  \right)
\right|_{\frac{q_k(i)}{P_{\max}}}
\nonumber \\
&&\!\!\!\!\!\!\!\!\!\!\! \! \! \!
+
\sum_{ k \in \mathcal K } \left. - \frac {g_k (i) \bar q_k (i)}{ \Gamma \sigma^2}\!  \left(\!
\frac{1} { W(\! - \! \text e ^{-\! 1 \! -\!  (\ln 2) ( \bar R_k ^{\text {UL} } ) ^{\alpha } \phi (i) } ) } \! +\! 1  \! \right)\!
\right|_0 \! =\! 1,
\end{eqnarray}
which can be solved by using the bisection method.

While given ($\pmb m, {\pmb n}$), the optimal transmission power and split power ($\pmb q ^*, \bar {\pmb q} ^*, \pmb v^*$) to problem (\ref{eem3max0}) is given by
{\color{myc2}{
\begin{equation}\label{Optalphamaxeq3}
q_k^* (i) = \left\{
\begin{array}{ll}
 \! \! v_k^* (i) & \text{if} \; \frac{\mu} { M } > \zeta \nu_k\sum _{l \in \mathcal K} g_l (i)  \\
\!\! P_{\max} m_k(i) & \text{if} \; \frac{\mu} { M } \leq  \zeta \nu_k\sum _{l \in \mathcal K} g_l (i)
\end{array}
\right.,
\end{equation}
}}
\begin{eqnarray}\label{Optalphamaxeq3_2}
\bar q_k ^* (i) =&&\!\!\!\!\!\!\!\!\!\!\! \!   \Bigg[ \frac{ n_k (i) }
{(\ln 2) ( \bar R_k ^{\text{UL}} ) ^{ \alpha } \nu_k \left( 1 - {\color{myc2}{\zeta_0}} \sum_{ l \in \mathcal K \setminus \{k\}  } g_{kl} (i) \right) }
\nonumber\\&&\!\!\!\!\!\!\!\!\!\!\! \!
 -\left.
\frac{  \Gamma \sigma^2 n_k(i)} {g_{k}(i)}
\Bigg]\right|_0,
\end{eqnarray}
and
\begin{equation}\label{Optalphamaxeq1}
v_k ^* (i) = \left. \left[\frac{   m_k (i) }{ (\ln 2)(\bar R_k ^{\text {DL}} )^{\alpha} \zeta \nu_k   g_{k } (i) } - \frac{  \Gamma \sigma^2 m_k (i) } { g_k (i) }\right] \right|^ {P_{\max} m_k(i)} _0
\end{equation}
for all $i\in\mathcal M, k \in \mathcal K$.
\end{theorem}

\itshape \textbf{Proof:}  \upshape
Please refer to Appendix D.
 \hfill $\Box$
 
From (\ref{Optalphamaxeq3_2}) and (\ref{Optalphamaxeq1}) in Theorem 3,  the optimal split power $v_k ^* (i)$ for each user in the DL increases with its allocated time $m_k(i)$, and the optimal transmission power $\bar q_k ^* (i)$ for each user in the UL also increases with the allocated time $n_k(i)$.
Based on Theorem 3, an implementation of the common fairness protocol at the BS is outlined by Algorithm 2.
In Algorithm 2, the average rate $\bar R_k ^{ \text X}$ in epoch $i$ is updated online according to a simple iterative rule \cite{Pejoski2017WPCN},
\begin{equation}\label{Optalphamaxeq5}
 \bar R_k ^{\text X} =  \frac{i-1}{i} \bar R_k ^{\text X} + \frac {1} {i} \bar r_k ^{\text {X}} (i), \quad \forall \text X \in \{ \text{DL}, \text {UL} \}.
\end{equation}

\begin{algorithm}[h]
\caption{: {\color{myc1}{Common Fairness Protocol Implementation at the BS}} }
\begin{algorithmic}[1]
\STATE Initialize $\pmb q$, $\bar { \pmb q}$, $\pmb v$, $ \mu $, and $ \pmb \nu $ ; Set $ \{ \lambda_ k\}_{ k\in \mathcal K \cup \{0\}} $ and $i=1 $.
\REPEAT
\REPEAT
\STATE Obtain the optimal $\{m_{k} (i), n _k (i) \}_{\forall k \in \mathcal K}$ with fixed $\{ q_k (i), \bar q_{k} (i), v_{k} (i)\}_{\forall k \in \mathcal K}$.
\STATE Obtain the optimal $\{q_k (i), \bar q_{k} (i), v _k (i)\}_{\forall k \in \mathcal K}$ with fixed $\{m_{k} (i), n _k (i) \}_{\forall k \in \mathcal K}$.
\UNTIL objective function (\ref{eem3max0}a) converges
\STATE Inform $\{ n _k (i), \bar q _k (i)/n_k(i), v_ k (i) /\bar q_k(i) \}_{\forall k \in \mathcal K}$ to all users, and broadcast RF energy at $p_k(i)=\frac{q_k (i)} { m _ k ( i ) }$ during $m_k (i)$, $\forall k \in \mathcal K$.
\STATE Update $\mu= \mu + \lambda_0  \left( \frac{1} i \sum_{t=1 } ^i \sum_{k \in \mathcal K} q_k(t) - P_{\text {avg  } } \right) $.
\STATE Update  $\nu_k= \nu_k +   \lambda_k    \left( \frac {1} i \sum_ {t=1} ^i \bar q_k (i) -  \frac {1} i  \sum_ {t=1} ^i \bar E_k(t)  \right) $, $\forall k \in \mathcal K$.
\STATE Update  $ \bar R_k ^{\text X} =  \frac{i-1}{i} \bar R_k ^{\text X} + \frac {1} {i} \bar r_k ^{\text {X}} (i) $, $\forall \text X \in \{ \text{DL}, \text {UL} \}, k \in \mathcal K$.
\STATE Set $i=i+1$.
\UNTIL $i \leq M$
\end{algorithmic}
\end{algorithm}

{\color{myc1}{
Note that the convergence of Algorithm 2 can be proved by using the same method in Section III-A.
The proof of convergence of Algorithm 2 is thus omitted.
}
}

\subsection{Max-Min Fairness with $\alpha=+\infty$}
According to \cite[Lemma~3]{879343}, problem (\ref{eem3max0}) with $\alpha=+\infty$ equivalently represents the following max-min problem\footnote{In problem (\ref{fto3max1}), the rates of both DL and UL are optimized as the same according to Lemma 2.
For the rates of DL and UL with different rate requirements, one can simply change the objective function (\ref{fto3max1}a) with
$\mathop{\max}_{ \pmb{m}, \pmb n,\pmb{q}, \bar {\pmb q}, \pmb v }\min_{k\in\mathcal K, \text X \in \{{\text{DL}}, {\text{UL}}\}}    \frac{\bar R_k^{\text{X}}}{\xi_k^{\text{X}}}$, where $ \xi_k ^{\text {X} } $ represents the rate requirements of user $k$ in the (X=)DL or (X=)UL.
With the modified objective function, the proposed method in this subsection is also valid.
}:
\begin{subequations}\label{fto3max1}
\begin{align}
\!\!\mathop{\max}_{ \pmb{m}, \pmb n,\pmb{q}, \bar {\pmb q}, \pmb v }\quad\!\!\!
&\min_{k\in\mathcal K, \text X \in \{{\text{DL}}, {\text{UL}}\}}    \bar R_k^{ \text{X} }
\\
\textrm{s.t.}\quad  \:\!\!\!
&(\ref{eem3max0}b), (\ref{eem3max0}c), (\ref{eem3max0}d), (\ref{eem3eq1}), (\ref{eem3max0}f).
\end{align}
\end{subequations}
By introducing an auxiliary variable, problem (\ref{fto3max1}) is equivalent to
\begin{subequations}\label{fto3max2}
\begin{align}
\mathop{\max}_{ \pmb{m}, \pmb n,\pmb{q}, \bar {\pmb q}, \pmb v }\qquad
&
\theta
\\
\textrm{s.t.}\qquad\:\:
&   \bar R_k^{\text{DL}}\geq \theta, \quad \forall k \in \mathcal K\\
&   \bar R_k^{\text{UL}}\geq \theta, \quad \forall k \in \mathcal K\\
&(\ref{eem3max0}b), (\ref{eem3max0}c), (\ref{eem3max0}d), (\ref{eem3eq1}), (\ref{eem3max0}f).
\end{align}
\end{subequations}

To solve problem (\ref{fto3max2}), we have the following lemma.
\begin{lemma}\label{faircond1}
Problem (\ref{fto3max2}) is designed to guarantee the rate of the users with the worse channel conditions.
The optimal ($\pmb m^*, \pmb n^*, \pmb q^*, \bar {\pmb q }^*, \pmb v^*, \theta^*$) of problem (\ref{fto3max2}) satisfies
\begin{equation}\label{fto3eq0}
(\bar R_k ^{\text{DL}})^*=
(\bar R_k ^{\text{UL}})^*= \theta^*,\quad \forall k \in \mathcal K.
\end{equation}
\end{lemma}

Lemma \ref{faircond1} can be easily proved by the contradiction method through the facts that $\bar R_k^{\text{DL}}$ is a monotonically increasing function of $(\pmb m, \pmb v)$ according to Appendix B and $\bar R_k^{\text{UL}}$ is a monotonically increasing function of ($\pmb n, \bar { \pmb q}$).

Although problem (\ref{fto3max2}) is convex based on the proof of Theorem 1, it is still hard to obtain the optimal solution in closed form.
In the following, we devise an alternating low-complexity method to solve problem (\ref{fto3max2}).

\begin{theorem}
Given ($\pmb q, \bar {\pmb q}, \pmb v$), the optimal time sharing factor to problem (\ref{fto3max2}), denoted by ($\pmb m^*, \pmb n^*$), equals
\begin{equation}\label{fto3eq2}
m_k^* (i)  = \left. - \frac {g_k (i) v_k (i)}{ \Gamma \sigma^2} \left(
\frac{1} { W(- \text e ^{-1 - (\ln 2)   \phi (i) /\psi_{1k} } ) } +1  \right)
\right|_{\frac{q_k(i)}{P_{\max}}},
\end{equation}
and
\begin{equation}\label{fto3eq2_2}
n_k^* (i)  = \left. - \frac {g_k (i) \bar q_k (i)}{ \Gamma \sigma^2} \left(
\frac{1} { W(- \text e ^{-1 - (\ln 2)   ^{\alpha } \phi (i) /\psi_{2k} } ) } +1  \right)
\right|_0
\end{equation}
for all $i\in\mathcal M, k \in \mathcal K$,
where $ \pmb \psi_1= \{\psi_{1k}\}_{k \in \mathcal K} $  and   $ \pmb \psi _2 = \{\psi_{2k}\}_{k \in \mathcal K} $ are respectively nonnegative Lagrangian multipliers associated with (\ref{fto3max2}b) and (\ref{fto3max2}c), and $\phi(i)$ is the root of the following equation
\begin{eqnarray}\label{fto3eq2_3}
&&\!\!\!\!\!\!\!\!\!\!\!\!\!\!\sum_{ k \in \mathcal K }
\left. - \frac {g_k (i) v_k (i)}{ \Gamma \sigma^2} \left(
\frac{1} { W(- \text e ^{-1 - (\ln 2)    \phi (i) / \psi_{1k} } ) } +1  \right)
\right|_{\frac{q_k(i)}{P_{\max}}}
\nonumber \\
&&\!\!\!\!\!\!\!\!\!\!\!\!\!\!
+
\sum_{ k \in \mathcal K } \left. - \frac {g_k (i) \bar q_k (i)}{ \Gamma \sigma^2} \left(
\frac{1} { W(- \text e ^{-1 - (\ln 2)   \phi (i) / \psi_{2k} } ) } +1  \right)
\right|_0 =1,\nonumber\\
\end{eqnarray}
which can be solved by using the bisection method.

While given ($\pmb m,  {\pmb n}$), the optimal transmission power and split power ($\pmb q^*, \bar {\pmb q} ^*, \pmb v^*$) to problem (\ref{fto3max2}) is
{\color{myc2}{
\begin{equation}\label{fto3eq3}
q_k^* (i) = \left\{
\begin{array}{ll}
 \! \! v_k^* (i) & \text{if} \; \frac{\mu} { M } > \zeta \nu_k\sum _{l \in \mathcal K} g_l (i)  \\
\!\! P_{\max} m_k(i) & \text{if} \; \frac{\mu} { M } \leq  \zeta \nu_k\sum _{l \in \mathcal K} g_l (i)
\end{array}
\right.,
\end{equation}
}}
\begin{equation}\label{fto3eq3_2}
\bar q_k ^* (i) \!=\!\!\left.\!\left[ \!\frac{ \psi_{2k} n_k (i) }
{(\ln 2) (   \nu_k \left( 1 -{\color{myc2}{ \zeta_0}} \sum_{ l \in \mathcal K \setminus \{k\}  } g_{kl} (i) \right) } -
\frac{\Gamma \sigma^2 n_k(i)} {g_k(i)}
\!\right]\!\right|_0\!\!\!, 
\end{equation}
and
\begin{equation}\label{fto3eq1}
v_k ^* (i) = \left.\left[ \frac{ \psi_{1k}  m_k (i) }{ (\ln 2)  \zeta \nu_k   g_{k } (i) } - \frac{  \Gamma \sigma^2 m_k (i) } { g_k (i) } \right] \right|^ {P_{\max}m_k (i)} _0
\end{equation}
for all $i\in\mathcal M, k \in \mathcal K$.
\end{theorem}

\itshape \textbf{Proof:}  \upshape
Please refer to Appendix E.
 \hfill $\Box$

Practical implementation of the max-min fairness protocol at the BS is similar to Algorithm~2.
The difference is that max-min fairness requires the updated information of $\pmb \psi_1$ and $\pmb \psi_2$.
Using the stochastic gradient descent method \cite{boyd2004ee364a} in each epoch, we have
\begin{equation}\label{fto3eq5}
\psi_{1k}= \psi_{1k} +   \lambda_{K+k}    \left(    \min_ {k\in\mathcal K, \text X \in \{{\text{DL}}, {\text{UL}}\}} \bar R_k^{\text{X}} - \bar R_k^{\text {DL } } \right),
\end{equation}
\begin{equation}\label{fto3eq5_1}
\psi_{2k}= \psi_{2k} +   \lambda_{2K+k}    \left(    \min_ {k\in\mathcal K, \text X \in \{{\text{DL}}, {\text{UL}}\}} \bar R_k^{\text{X}} - \bar R_k^{\text { UL } } \right),
\end{equation}
and
\begin{equation}\label{fto3eq5_2}
\psi_{1k}= \frac {\psi_{1k}} {\sum_{k \in \mathcal K} ( \psi_{1k}+ \psi_{2k} ) } , \psi_{2k}= \frac { \psi_{2k} } {\sum_{k \in \mathcal K} ( \psi_{1k}+ \psi_{2k} ) }
\end{equation}
for all $k\in\mathcal K$,
where $ \{ \lambda_{K+ k},  \lambda_{2K+ k}\}_{ k\in \mathcal K} $ are some step sizes.
Equation (\ref{fto3eq5_2}) follows from the optimal condition given in (\ref{appenEkkt1}a) and (\ref{appenEkkt2}a) in Appendix E.



\subsection{Complexity Analysis}
Considering that the dimension of the variables in problem (\ref{eem3max0}) is $5KM$, the complexity of solving problem (\ref{eem3max0}) by using the standard
IPM is $\mathcal O(L_{\text{IP}}K^3M^3)$ \cite[Pages 487, 569]{boyd2004convex}, where $L_{\text{IP}}$ is the number of iterations.

For the proposed zero fairness protocol implementation (ZFPI) in Algorithm 1, the complexity of solving $\{m_k(i),  n_k(i),  q_k(i), \bar { q }_k(i), v_k(i) \}_{ k \in \mathcal K } $ in epoch $i$ is $\mathcal O(L_{\text{ZF}} K)$, where $L_{\text{ZF}}$ is the number of iterations for objective function (\ref{eem3max1}a) to converge.
As a result, the total complexity of ZFPI is $\mathcal O(L_{\text{ZF}} KM)$.

For the proposed common fairness protocol implementation (CFPI) in Algorithm 2,
the major complexity lies in the computation of time factor $\{m_k(i), n_k(i)\}_{k \in \mathcal K}$.
According to (\ref{Optalphamaxeq2}) and (\ref{Optalphamaxeq2_2}), the complexity of calculating $\{m_k(i), n_k(i)\}_{k \in \mathcal K}$ with given $\phi(i)$ is $\mathcal O(K\log_2(1/\epsilon_1))$,
where $\mathcal O(\log_2(1/\epsilon_1))$ is the complexity of calculating the Lambert-W function with the accuracy of $\epsilon_1$.
To calculate $\phi(i)$, the complexity is $\mathcal O(\log_2(1/\epsilon_2))$, where $\mathcal O(\log_2(1/\epsilon_2))$ is the complexity of solving (\ref{Optalphamaxeq2_3}) by using the bisection method with the accuracy of $\epsilon_2$.
Thus, the total complexity of computing $\{m_k(i), n_k(i)\}_{k \in \mathcal K}$ is $\mathcal O(K\log_2(1/\epsilon_1)\log_2(1/\epsilon_2))$.
As a result, the total complexity of CFPI is
$\mathcal O(L_{\text{CF}} KM\log_2(1/\epsilon_1)\log_2(1/\epsilon_2))$,
where $L_{\text{CF}}$ is the total number of required iterations for the convergence of (\ref{eem3max0}a).
For convenience, the accuracies are set to the same value, i.e., $\epsilon_1=\epsilon_2=\epsilon$.
The total complexity of the CFPI can be simplified as
$\mathcal O(L_{\text{CF}} KM\log_2^2(1/\epsilon))$.
With the same analysis, the total complexity of the max-min fairness protocol implementation (MFPI) is
$\mathcal O(L_{\text{MF}} KM\log_2^2(1/\epsilon))$, where $L_{\text{MF}}$ is the total number of required iterations for the convergence of (\ref{fto3max2}a).
Compared with the IPM, we can find that the proposed ZFPI and CFPI have a lower order of complexity.

\section{Numerical Results}

In this section, we evaluate the performance of the proposed algorithms.
Both the DL and UL channel power gains are modeled as $g_k=10^{-3}\rho_k^2D_k^{-\chi}$ \cite{6678102}, $k=1, \cdots, K$, where $\rho_k$ represents the short-term fading, which indicates that $\rho_k^2$ is an exponentially distributed random variable with unit mean,
and $D_k$ (in meter) is the distance between user $k$ and the BS.
Note that according to the above channel model, a 30 dB average signal power attenuation is assumed at a reference distance of 1 m, and the pathloss exponent is set to $\chi=3$.
There are a total number of $K$, ranging from 2 to 10, users uniformly distributed in a square area 10 m $\times$ 10 m.
For each user, the energy harvesting efficiency is assumed to be $\zeta=0.5$.
The noise power is $\sigma^2=-104$ dBm and we set $\Gamma=9.8$ dB assuming that an uncoded quadrature amplitude modulation (QAM) is employed \cite{goldsmith2005wireless}.
The number of epochs is st as $M=10^3$.
Unless specified otherwise, the system parameters are set as {\color{myc2}{$\zeta_0=0.5$}}, $P_{\max}=5$ W, and $K=10$.

\subsection{Sum Rate and Fairness}
\begin{figure}
\centering
\includegraphics[width=3.2in]{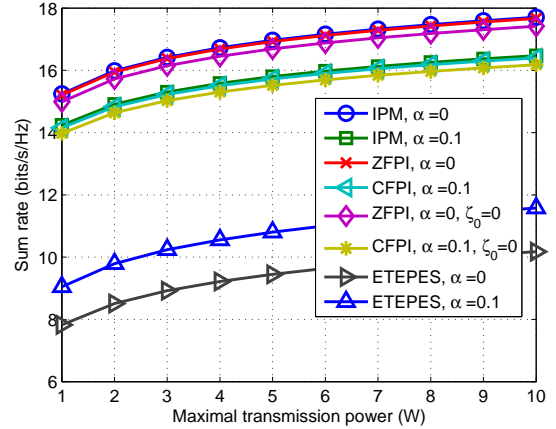}
\caption{Sum rate vs. maximal transmission power of the BS.\label{fig13}}
\end{figure}

In this subsection, we provide the sum rate and fairness performance
with various values of $\alpha$.
We compare the proposed ZFPI and CFPI schemes with the IPM scheme to solve problem (\ref{eem3max0}) by using the matlab cvx toolbox \cite{grant2009cvx}, and the equal time allocation, equal transmission power and equal power-splitting factor (ETEPES) scheme, i.e.,
$m_k(i)=n_k(i)=\frac 1 {2K}$, $q_k(i)=q_j(i)$, $\bar q_k(i)=\bar q_j (i)$, $v_k(i)=q_k(i)/2$, $\forall k, j \in \mathcal K, i \in \mathcal M$.
Fig.~\ref{fig13} shows the {\color{myc1}{sum rate}} versus maximal transmission power of the BS.
From Fig.~\ref{fig13}, the sum rate achieved by the ZFPI or CFPI is almost the same as the IPM, while the complexity of the ZFPI or CFPI is much lower compared to IPM  according to Section III-D.
It is also observed that the proposed sum rate is superior over the EFEPES, which shows the effectiveness of the proposed algorithm.
{\color{myc2}{According to Fig.~\ref{fig13}, the proposed sum rate with $\zeta_0=0$ is slightly lower than that with $\zeta_0=0.5$, which is, as expected, due to the fact that the users cannot harvest the energy from the other users for the case $\zeta_0=0$.}}
Note that the presented values in Fig.~\ref{fig13} seem too high due to the following two reasons.
The first reason is that sum rate includes both DL and UL rates.
The second reason is that the channel gains between the BS and users are relatively higher than in conventional cellular communication networks due to low distance between the BS and the users in WPCNs.

For a given $\alpha$, we utilize Jain's index to evaluate the degree of fairness.
Jain's index is defined as
\begin{equation}
\text{Jain's index} = \frac{\left(\sum_{k \in \mathcal K}(\bar R_{k}^{\text{DL}}+\bar R_{k}^{\text{UL}})\right)^2}
{2K\sum_{k \in \mathcal K}((\bar R_{k}^{\text{DL}})^2+(\bar R_{k}^{\text{UL}})^2)},
\end{equation}
which is bounded in $[1/(2K), 1]$ \cite{jain1984quantitative}.
{\color{myc1}{Within this interval, Jain's index = 1/$(2K)$ corresponds to the least fair allocation in which only one rate receives a non-zero benefit, and Jain's
index = 1 corresponds to the fairest allocation in which all UL and DL rates receive the same benefit \cite{6816520}.
Therefore, the Jain's index quantitatively represents the fairness between users, and a relatively large Jain's index indicates that the differences of different user rates are not distinctive.
As a result, a larger Jain's index is more fair \cite{6816520,5461911}.
}}

\begin{figure}
\centering
\includegraphics[width=3.2in]{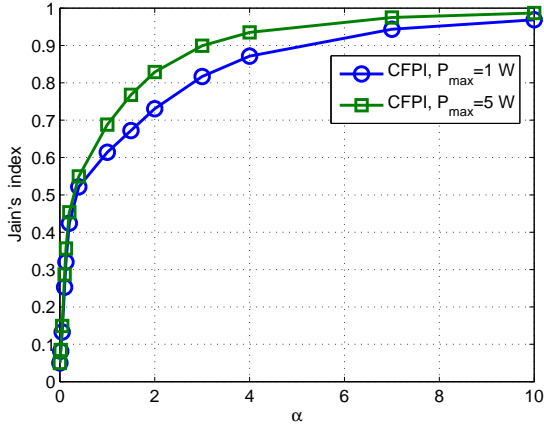}
\caption{Jain's index vs. $\alpha$.\label{fig15}}
\end{figure}

\begin{figure}
\centering
\includegraphics[width=3.2in]{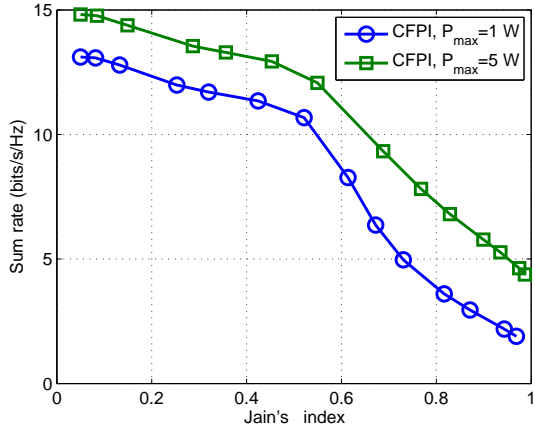}
\caption{Sum rate vs. Jain's index.\label{fig16}}
\end{figure}

Fig.~\ref{fig15} shows Jain's index versus $\alpha$ under various maximal transmission power constraints of the BS.
Jain's index increases monotonically with the value of $\alpha$, i.e.,
if we increase $\alpha$, a more fair time and power allocation can be obtained.
The increasing speed is rapid at low-value region of $\alpha$,
while the speed is slow at high-value region of $\alpha$.
The tradeoff curve between sum rate and Jain's fairness is presented in Fig.~\ref{fig16}.
Combining Fig.~\ref{fig15} and Fig.~\ref{fig16}, it is observed that the largest sum rate is obtained
at $\alpha=0$, while the solution is least fairness.
The largest Jain's index 1 is obtained when $\alpha=+\infty$, which yields the smallest sum rate.

\begin{figure}
\centering
\includegraphics[width=3.2in]{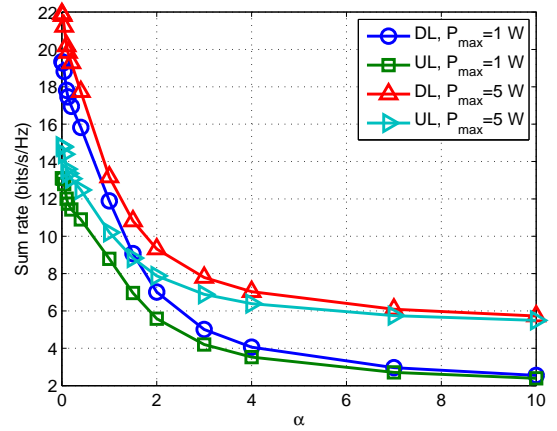}
\caption{{\color{myc1}{Sum rate for DL and UL vs. $\alpha$.\label{fig17}}}}
\end{figure}

{\color{myc1}{
The sum rate for DL and UL of the proposed CFPI method versus $\alpha$ under various maximal transmission power constraints of the BS is presented in Fig.~\ref{fig17}.
According to Fig.~\ref{fig17}, the sum rate for both DL and UL decreases with the value of $\alpha$.
It is also observed that the gap of sum rate between DL and UL also decreases with the value of $\alpha$, which implies that the fairness between DL and UL increases with $\alpha$.}}

\subsection{Zero Fairness}
In this subsection, we evaluate the sum rate performance for the zero fairness case,
i.e., $\alpha=0$.
We compare the proposed ZFPI with the ETEPES scheme, the
sum throughput scheme where only DL wireless energy transfer is considered (ST-DWET) \cite{6678102},
and the proposed MFPI in Section III-C.

\begin{figure}
\centering
\includegraphics[width=3.2in]{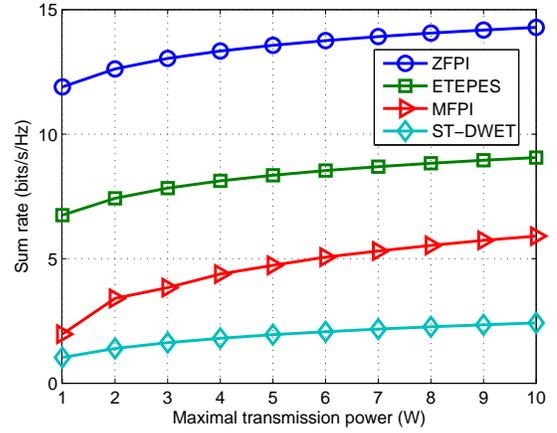}
\caption{Sum rate vs. maximal transmission power of the BS.\label{fig3}}
\end{figure}

Fig.~\ref{fig3} shows sum rate versus maximal transmission power of the BS.
It is observed that ZFPI, ETEPES and MFPI all outperforms ST-DWET.
This is because ZFPI, ETEPES and MFPI all considers simultaneous wireless information and power transfer in the DL and {\color{myc1}{user can harvest additional energy when other users transmit UL information}}, while ST-DWET only allows wireless energy transfer in the DL.

\begin{figure}
\centering
\includegraphics[width=3.2in]{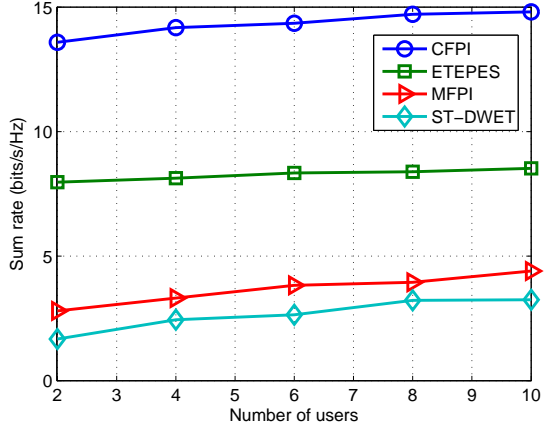}
\caption{Sum rate vs. number of users. \label{fig4}}
\end{figure}

The sum rate comparison for different number of users is shown in Fig.~\ref{fig4}.
It is observed that sum rate increases with the number of users for all schemes.
Based on Fig.~\ref{fig3} to Fig.~\ref{fig4}, it is observed that the proposed ZFPI achieves the best sum rate among four schemes. 

\subsection{Max-Min Fairness}
In this subsection, we evaluate the fairness value for the max-min fairness case, i.e., $\alpha=+\infty$.
We compare the proposed ZFPI scheme in Section III-C with the ETEPES scheme, the equal time allocation, equal transmission power and optimized split power (ETEPOS) scheme (i.e., $m_k(i)=n_k(i)=\frac 1 {2K}$, $q_k(i)=q_j(i)$, $\bar q_k(i)=\bar q_j(i)$, $\forall k, j \in \mathcal K, i \in \mathcal M$, optimal $\pmb v$ is obtained from (\ref{fto3eq1})),
and
the optimized time and transmission power allocation and equal power-splitting factor (OTOPES) scheme (i.e., $v_k(i)=\frac{q_k(i)}{2}$, $\forall k \in \mathcal K, i \in \mathcal M$, optimal ($\pmb m, \pmb n, \pmb q, \bar {\pmb q}$) is obtained from Theorem 4).

\begin{figure}
\centering
\includegraphics[width=3.2in]{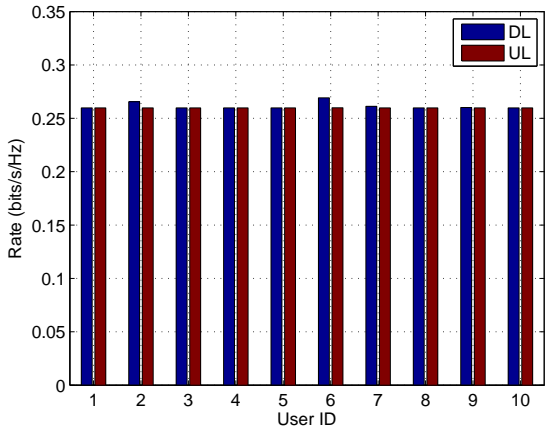}
\caption{{\color{myc1}{Rate distribution of each user in both DL and UL.\label{fig12}}}}
\end{figure}

{\color{myc1}{
Fig. \ref{fig12} shows the rate distribution of each user in both DL and UL of the proposed MFPI mrthod.
From Fig.~\ref{fig12}, the rates for each user's DL and UL are almost the same, which verifies the fairness achieved by the proposed MFPI.}}

\begin{figure}
\centering
\includegraphics[width=3.2in]{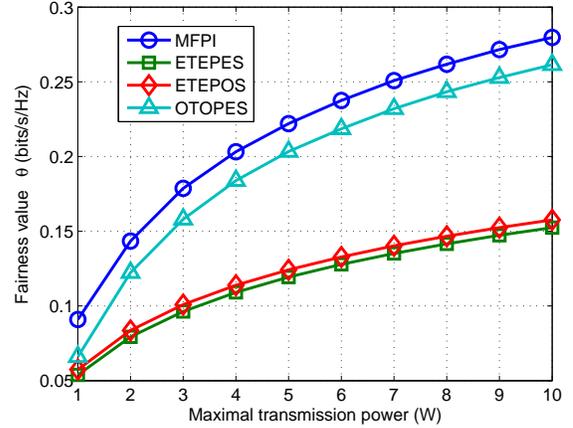}
\caption{Fairness value vs. maximal transmission power of the BS. \label{fig6}}
\end{figure}

In Fig.~\ref{fig6}, we illustrate the fairness value $\theta$ in (\ref{fto3max2}) versus maximal transmission power of the BS.
{\color{myc1}{As shown in Fig.~\ref{fig6}, the proposed MFPI achieves better performance than OTOPES, which verifies that the system performance can be further improved with the optimization of split power.
The ETEPES yields the least fairness value, since neither time allocation nor power control is optimized.
It is observed that the OTOPES outperforms the ETEPOS, which demonstrates that the optimization of time allocation and transmission power dominates the optimization of split power.
This is because that the rates in both DL and UL grow almost linearly with time allocation vector, while grows logarithmically with split power according to (\ref{eem3max0eq1}) and (\ref{eem3max0eq1_2}).}}

\begin{figure}
\centering
\includegraphics[width=3.2in]{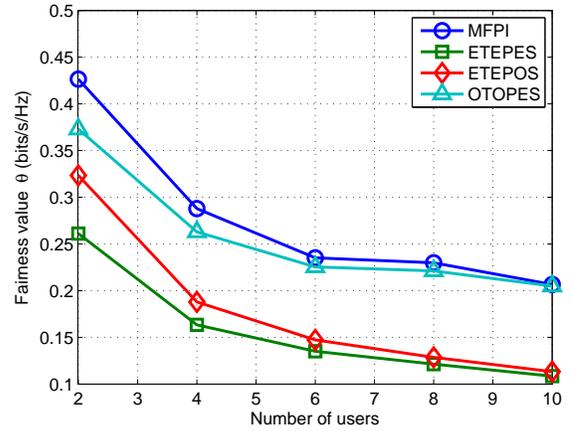}
\caption{Fairness value vs. number of users. \label{fig9}}
\end{figure}

The fairness value comparison for different numbers of users is shown in Fig.~\ref{fig9}.
According to Fig.~\ref{fig9}, it is observed that the fairness value decreases with the increase of number of users. 
This is because that high number of users can degrade the system performance.  

\section{Conclusion}
In this paper, we have investigated the sum $\alpha$-fair utility maximization for a WPCN with simultaneous wireless information and power transfer in both DL and UL.
The general sum $\alpha$-fair utility maximization has been cast to a convex problem.
For zero fairness, the optimal time allocation is proportional to the available transmission power, and it is optimal for at most one user in each epoch to transmit with positive power in the UL.
For max-min fairness, the optimization of time allocation and transmission power with equal power-splitting factor is superior over the optimization of split power with equal time allocation and transmission power.
Simulation results show that the sum rate of the proposed scheme outperforms the conventional WPCN with only energy transfer in the DL and information transfer in the UL.
In a WPCN with multi-antenna BS,
however, it is generally challenging to obtain the globally optimal time and power allocation.
A direct extension can be made resorting to an alternating method, i.e., iteratively optimizing DL transmission and receiving beamforming, and updating time and power allocation strategies.
{\color{myc2}{The  synchronization of the DL and UL slots and the non-ideal energy harvesting characteristics of the users are left for our future work.}}

\appendices
\section{Proof of Theorem 1}
\setcounter{equation}{0}
\renewcommand{\theequation}{\thesection.\arabic{equation}}
Obviously, the function $f(v_k(i))\triangleq\log_2 \left(1+\frac{g_{k}(i)v_{k}(i)}{\Gamma\sigma^2}\right)$ is concave w.r.t. $v_{k}(i)$.
According to the property of perspective function \cite[Section~3.2.6]{boyd2004convex},
$\bar r_k^{\text{DL}}(i)=m_k(i)f(\frac{v_k(i)}{m_k(i)})=m_{k}(i)$ $\log_2 \left(1+\frac{g_{k}(i)v_{k}(i)}{\Gamma \sigma^2m_{k}(i)}\right)$ is concave w.r.t. ($m_k(i), v_{k}(i)$).
From (\ref{sys2eq7}), we have
\begin{equation} \label{lemma1eq1}
U'_{\alpha}(x)=x^{-\alpha}\geq0, \quad\forall \alpha\geq0, x\geq0,
\end{equation}
and
\begin{equation}\label{lemma1eq2}
U''_{\alpha}(x)=-x^{-\alpha-1}\leq0, \quad\forall \alpha\geq0, x\geq0,
\end{equation}
which demonstrates that $U_{\alpha}(x)$ is concave and nondecreasing.
Based on the property of scalar composition \cite[Section~3.2.4]{boyd2004convex}, $U_{\alpha}\left(\frac{1} M \sum_{i\in\mathcal M} \bar r_k^{\text{DL}}(i)\right)$ is concave w.r.t. ($\pmb m, \pmb v$).
Analogously, we can prove that both
$\bar r_k^{\text{UL}} (i)=n_k(i) \log_2\left(1+\frac{g_{k}(i)\bar q_{k}(i)}{\Gamma \sigma^2 n_{k}(i)}\right)$ and $U_{\alpha}\left(\frac{1} M \sum_{i\in\mathcal M} \bar r_k^{\text{UL}}(i)\right)$ are concave w.r.t. ($\pmb n, \bar {\pmb q}$).
Since $\sum_{k\in \mathcal K} U_{\alpha}\left(\frac{1} M \sum_{i\in\mathcal M} \bar r_k^{\text{DL}}(i)\right) +  \sum_{k\in \mathcal K}U_{\alpha}\left(\frac{1} M \sum_{i\in\mathcal M} \bar r_k^{\text{DL}}(i)\right)$ is a nonnegative weighted sum of concave functions.
Hence, the objective function of problem (\ref{eem3max0}) is concave w.r.t. $(\pmb m, \pmb n,  \bar{\pmb q}, \pmb v)$ \cite[Section~3.2.1]{boyd2004convex}.

\section{Proof of Lemma 1}
\setcounter{equation}{0}
\renewcommand{\theequation}{\thesection.\arabic{equation}}
Assume that  $\sum_{k\in \mathcal K}m_{k}^*(i)+n_k^*(i)<1$.
The first-order derivative of $\bar r_{k}^{\text{DL}}(i)$ w.r.t. $m_k(i)$ is
\begin{eqnarray}\label{lemma2firsetder1}
\frac{\partial \bar r_{k}^{\text{DL}}(i)}
{\partial m_{k}(i)}
 =&&\!\!\!\!\!\!\!\!\!
  \frac{1}{\ln2}
 \ln\left(1 + \frac{g_{k}(i) v_{k}(i)}{\Gamma\sigma^2m_{k}(i)}\right)
 \nonumber\\ &&\!\!\!\!\!\!\!\!\!
 - \frac{g_{k}(i)v_{k}(i)}{(\ln 2)({g_{k}(i)v_{k}(i)} + \Gamma\sigma^2m_{k}(i))}.
\end{eqnarray}
Define function
\begin{equation}\label{lemma2deffunction}
h(x)=\frac{ \ln(1+x)}{\ln2} -\frac{x}{(\ln2 )(1+x)}, \quad \forall x \geq 0.
\end{equation}
From (\ref{lemma2firsetder1}), we have $\frac{\partial \bar r_{k}^{\text{DL}}(i)}
{\partial m_{k}(i)}= h\left(\frac{g_{k}(i) v_{k}(i)}{\Gamma\sigma^2m_{k}(i)}\right)$.
Since
\begin{equation}\label{lemma2functiong}
h'(x)=\frac{(\ln 2)x}{(x+1)^2}> 0, \quad \forall x>0,
\end{equation}
we obtain $h(x)> h(0)=0$.
Then, $\frac{\partial \bar r_{k}^{\text{DL}}(i)}
{\partial m_{k}(i)} \geq 0$ and $\bar r_{k}^{\text{DL}}(i)$ is increasing for $m_{k}(i) \geq 0$.
According to (\ref{lemma1eq1}), $U_{\alpha}\left(\frac 1 M \sum_{i \in \mathcal M} \bar r_k^{\text{DL}}(i)\right)$ is also increasing for $m_{k}(i) \geq 0$.
Thus, objective function (\ref{eem3max0}a) can always be improved by increasing $m_{k}^*(i)$, contradicting that solution ($\pmb m^*, \pmb n^*$) is optimal.

\section{Proof of Theorem 2}
\setcounter{equation}{0}
\renewcommand{\theequation}{\thesection.\arabic{equation}}
With fixed $(\pmb q, \pmb v)$, the Lagrange function of problem (\ref{eem3max1}) can be written by
\begin{eqnarray}
&&\!\!\!\!\!\!\!\!\!\!\!\mathcal L_1
(\pmb{m}, \pmb n, \bar { \pmb q }, \pmb \phi, \pmb \nu, \pmb \eta_0, \pmb \eta_1, \pmb \eta_2, \pmb \eta_3)
=
\nonumber\\&&\!\!\!\!\!\!\!\!\!\!\!
\frac{1} M \sum_{i\in\mathcal M} \sum_{k\in \mathcal K} m_k (i) \log_2 \left(1+\frac{g_k(i) v_k(i) }{\Gamma \sigma^2 m_k(i)} \right)
\nonumber\\&&\!\!\!\!\!\!\!\!\!\!\!
+  \frac 1 M \sum_{i\in\mathcal M}\sum_{k\in \mathcal K} n_k (i) \log_2\left(1+
\frac{
 g_k (i) \bar q_k (i)}
{\Gamma \sigma^2 n_k (i) }\right)
\nonumber\\&&\!\!\!\!\!\!\!\!\!\!\!
+ \sum_{i\in\mathcal M} \phi (i) \left( 1- \sum_{k\in \mathcal K} ( m_k(i) + n_k (i) )
\right)
\nonumber\\&&\!\!\!\!\!\!\!\!\!\!\!
+ {\color{myc2}{\sum_{i\in\mathcal M}\sum_{k \in \mathcal K} \eta_{0k} (i) \left( P_{\max} m_k (i) - q_k(i)
\right)}}
\nonumber\\&&\!\!\!\!\!\!\!\!\!\!\!
+\sum _{ i \in \mathcal M } \sum_{k \in \mathcal K} \nu_k \left(  \zeta g_k(i) \sum_{l\in\mathcal K } q_l (i) -  \zeta g_k (i) v_k (i)
\right.
\nonumber\\&&\!\!\!\!\!\!\!\!\!\!\! \left.
+
   {\color{myc2}{\zeta_0}} \sum_{l=1} ^{k-1} g_{lk} (i) \bar q_l (i)+
   {\color{myc2}{\zeta_0}} \sum_{l=k+1}^K g_{lk} (i-1)  \bar q_l (i-1)- \bar q_k (i)
      \right)
      \nonumber\\&&\!\!\!\!\!\!\!\!\!\!\!  +  \sum_{ i\in \mathcal M }   \sum _{k\in \mathcal K} \eta_{1k} (i) m _k (i)+  \sum_{ i\in \mathcal M }  \sum _{k\in \mathcal K}  \eta_{2k} (i) n _k (i)
 \nonumber\\&&\!\!\!\!\!\!\!\!\!\!\!
+ \sum_{ i\in \mathcal M } \sum _{k\in \mathcal K} \eta_{3k} (i) \bar q_k (i),
\end{eqnarray}
where $\pmb \phi = \{\phi (i) \}_{i\in\mathcal M}$, $\pmb \nu=\{\nu_k\}_{k \in \mathcal K}$, and $\pmb \eta_ j =\{\eta_{jk}(i) \}_{i \in\mathcal M, k \in \mathcal K}$, $j=0, 1, 2, 3$, are nonnegative Lagrangian multipliers associated with the corresponding constraints of problem (\ref{eem3max1}).
According to \cite{boyd2004convex}, the optimal solution of problem (\ref{eem3max1}) given ($\pmb q, \pmb v$) should satisfy:
\begin{subequations}\label{appentheorem1kkt1}
\begin{align}
&\frac { \partial \mathcal L_1}
{\partial m_k(i)}=
h\left(\frac{g_k (i) v_k (i)} { \Gamma \sigma^2 m_k (i) } \right)
 - \phi (i) +{\color{myc2}{\eta_{0k}(i)}}  +\eta_{1k}(i)= 0 \\
& \frac { \partial \mathcal L_1}
{\partial n_k(i)}
=h \left(\frac{g_k (i) \bar q_k (i)}{\Gamma\sigma^2 n_k(i)}\right)
 - \phi (i) + \eta_{2k}(i) = 0 \\
 &\frac{\partial \mathcal L_1}
{\partial \bar q_k (i)}
= \frac{ g_k (i) n_k (i) } { (\ln 2) ( \Gamma \sigma^2 n_k (i) + g_k (i) \bar q_k (i) ) }
\nonumber
\\&\qquad\qquad
 + \zeta \nu_k \sum_{ l \in \mathcal K \setminus \{k\}  } g_{kl} (i) - \nu_k + \eta_{ 3 k } (i) = 0,
\end{align}
\end{subequations}
for all $i \in \mathcal M$, $ k \in \mathcal K $,
where function $h(x)$ is defined in (\ref{lemma2deffunction}).

For the optimal solution to problem (\ref{eem3max1}), we observe that $m_k (i) =0 $ if and only if $q_k (i) =0$ and $n_k (i) =0$ if and only if $\bar q_k (i) =0$, as otherwise the sum rate (\ref{eem3max1}a) can be further increased.
Based on (\ref{appentheorem1kkt1}a) and the complementary slackness conditions, we have
\begin{equation}\label{appentheorem1kkt1_m_1}
m_k (i)=
\left. \frac{g_k (i) v_k (i) } { C(i) } \right|_{\frac{q_k(i)}{P_{\max}}},
\end{equation}
where $ { C(i) }=\frac{\Gamma \sigma^2} {h^{-1}(\phi(i))}$, $h^{-1}(x)$ is the inverse function of $h(x)$, and $a|_b=\max\{a,b\}$.
From (\ref{appentheorem1kkt1}b), $n_k(i)$ can be derived as
\begin{equation}\label{appentheorem1kkt1_m_3}
n_k (i) =\frac{ g_k (i) \bar q_k (i) } { C(i) }.
\end{equation}
Substituting (\ref{appentheorem1kkt1_m_1}) and (\ref{appentheorem1kkt1_m_3}) into (\ref{eem3max0}b) with equality from Lemma 1,
we have
\begin{equation}\label{appentheorem1kkt5}
\sum_{k\in \mathcal K} \left(
\left. \frac{g_k (i) v_k (i) } { C(i) } \right|_{\frac{q_k(i)}{P_{\max}}}+
 \frac{ g_k (i) \bar q_k (i) } { C(i) }\right)
 =1.
\end{equation}
Since the left term of equation (\ref{appentheorem1kkt5}) monotonically decreases with $C(i)$, the root $C(i)$ satisfying (\ref{appentheorem1kkt5}) with given $\bar {\pmb q}$ can be obtained via the bisection method.

Plugging (\ref{appentheorem1kkt1_m_3}) into (\ref{appentheorem1kkt1}c) yields
\begin{equation} \label{appentheorem1kkt5_6}
\eta_{ 3 k } (i) =  - \frac{ g_k (i)  } { (\ln 2) ( \Gamma \sigma^2 + C(i) ) } +  \nu_k\left( 1-{\color{myc2}{\zeta_0}} \sum_{ l \in \mathcal K \setminus \{k\}  } g_{kl} (i) \right).
\end{equation}
Combining (\ref{appentheorem1kkt5_6}) and $\eta_{3k} (i) \geq 0$, we get
\begin{equation} \label{appentheorem1kkt5_7}
C(i) \geq   \frac{ g_k (i)  } { (\ln 2)\nu_k\left( 1-{\color{myc2}{\zeta_0}} \sum_{ l \in \mathcal K \setminus \{k\}  } g_{kl} (i) \right) } - \Gamma \sigma^2 \triangleq D_k (i) .
\end{equation}
Assume that there exists one unique minimal value in sequence $\{D_1 (i) , \cdots, D_K (i) \}$,  and
$w (i) = \arg \min_ {k\in\mathcal K} D_ k (i) $.
When there exists more than one maximal values in sequence $\{D_1 (i) , \cdots, D_K (i) \}$, we can arbitrarily choose one index $w (i)$ with the maximal value $D_{w(i)}(i)$.
Based on (\ref{appentheorem1kkt5_6}) and (\ref{appentheorem1kkt5_7}), we have $ \eta_{ 3 k} (i) > \eta _{ 3 w(i) } (i) \geq  0$ and
\begin{equation} \label{appentheorem1kkt5_8}
\bar q_k (i) = 0,  \quad \forall k \in \mathcal K \setminus \{ w(i) \}.
\end{equation}
For nonnegative multiplier $ \eta _{3 w (i)} (i)$, two cases are considered.
\begin{enumerate}
  \item If $\eta _ {3w} (i) >0$, we have $\bar q_ {w(i)} (i) = 0 $ and
  $ \tilde C(i)$ satisfying $\sum_{k\in \mathcal K}
\left. \frac{g_k (i) v_k (i) } { \tilde C(i) } \right|_{\frac{q_k(i)}{P_{\max}}} =1$
  from  (\ref{appentheorem1kkt5}) and (\ref{appentheorem1kkt5_8}).
  Considering (\ref{appentheorem1kkt5_7}), we further have $\tilde C(i) > D_{ w(i) } (i) $.
  \item If $\eta _ { 3 w( i ) } (i) =0$, we have $C(i) = D_{ w(i) } (i)$.
  Based on (\ref{appentheorem1kkt5}) and (\ref{appentheorem1kkt5_8}), we can obtain $\bar q_ {w(i)} (i) = \frac{ D_{ w(i) } (i) - \sum _{ k \in \mathcal K } g_k (i) v_k (i)|_{ {q_k(i) D_{w(i)}}/{P_{\max}}}} {g_{w(i)} (i)} $, which indicates that $ \tilde C(i)\leq D_{ w(i) } (i) $.
\end{enumerate}
As a result, the optimal $\bar q _ k (i)$ is presented in (\ref{eem3eq3}) and (\ref{eem3eq2_2}).

Given ($ \pmb m, \pmb n , \bar {\pmb q}$), the Lagrange function of problem (\ref{eem3max1}) can be written by
\begin{eqnarray}
&& \!\!\!\!\!\!\!\!\! \mathcal L_2
(\pmb{q}, \pmb v, \pmb \nu, \mu, \pmb \eta_0, \pmb \eta_4, \pmb \eta_5)
=
\nonumber\\&& \!\!\!\!\!\!\!\!\!
\frac{1} M \sum_{i\in\mathcal M} \sum_{k\in \mathcal K} m_k (i) \log_2 \left(1+\frac{g_k(i) v_k(i) }{\Gamma \sigma^2 m_k(i)} \right)
\nonumber\\&& \!\!\!\!\!\!\!\!\!+ \sum_{k \in \mathcal K} \sum _{ i \in \mathcal M } \nu_k
 \left(  \zeta g_k(i) \sum_{l\in\mathcal K } q_l (i)
 -  \zeta g_k (i) v_k (i)
\right.
\nonumber\\&&\!\!\!\!\!\!\!\!\!\left.+
   {\color{myc2}{\zeta_0}} \sum_{l=1} ^{k-1} g_{lk} (i) \bar q_l (i)
+
   {\color{myc2}{\zeta_0}} \sum_{l=k+1}^K g_{lk} (i-1)  \bar q_l (i-1)- \bar q_k (i)
      \right)
 \nonumber\\&&\!\!\!\!\!\!\!\!\!
+  {\color{myc2}{\sum_{i\in\mathcal M}\sum_{k \in \mathcal K} \eta_{0k} (i) \left( P_{\max} m_k (i) - q_k(i)
\right)}}
 \nonumber\\&&\!\!\!\!\!\!\!\!\!+ \mu\left( P_{\text {avg} } - \frac 1 M \sum_{i\in\mathcal M } \sum_{k \in \mathcal K} q_k(i)  \right)  +  \sum_{ i\in \mathcal M }   \sum _{k\in \mathcal K} \eta _{5k} (i)  v_k  (i)
\nonumber\\&&\!\!\!\!\!\!\!\!\!
+
\sum_{ i\in \mathcal M }  \sum _{k\in \mathcal K}  \eta_{ 4k } (i)   (q _k (i) -v_k (i)),
\end{eqnarray}
where $\mu$, $\pmb \eta_4 = \{  \eta_{  4 k } (i) \}_{i\in\mathcal M, k \in \mathcal K} $ and $\pmb \eta_5 = \{  \eta_{  5k } (i) \}_{i\in\mathcal M, k \in \mathcal K} $ are nonnegative Lagrangian multipliers associated with the corresponding constraints of problem (\ref{eem3max1}).
According to \cite{boyd2004convex}, the optimal solution of problem (\ref{eem3max1}) with given ($\pmb m, \pmb n, \bar {\pmb q}$) should satisfy:
\begin{subequations}\label{appentheorem1kkt1optv}
\begin{align}
&\frac{\partial \mathcal L_2}
{\partial q_k (i)}
=- {\color{myc2}{\eta_{0k} (i)}} - \frac{\mu} { M } +\zeta \nu_k\sum _{l \in \mathcal K} g_l (i) + \eta_{ 4 k } (i) =0 \\
&\frac{\partial \mathcal L_2}
{v_k (i)}
= \frac{ g_k (i) m_k (i) } { (\ln 2) ( \Gamma \sigma^2 m_k (i) + g_k (i) v_k (i) ) } - \zeta \nu_k   g_{k } (i)\nonumber
 \\&\qquad\qquad +  \eta_{ 5k } (i) - \eta _{ 4k } (i) = 0,
\end{align}
\end{subequations}
for all $i \in \mathcal M$, $ k \in \mathcal K $.


{\color{myc2}{
According to (\ref{appentheorem1kkt1optv}a), we obtain
\begin{equation}\label{appentheorem1kkt5_1}
 \eta_{4k} (i)-\eta_{0k}(i) =  \frac{\mu} { M } - \zeta \nu_k\sum _{l \in \mathcal K} g_l (i).
\end{equation}
If $\frac{\mu} { M } - \zeta \nu_k\sum _{l \in \mathcal K} g_l (i)>0$, we have $\eta_{4k}(i)>0$ since $\eta_{0k}(i)\geq 0$ and $q_k(i)=v_k(i)$ according to the complementary slackness conditions.
If $\frac{\mu} { M } - \zeta \nu_k\sum _{l \in \mathcal K} g_l (i)<0$, we have $\eta_{0k}(i)>0$ since $\eta_{4k}(i)\geq 0$ and $q_k(i)=P_{\max}m_k(i)$ according to the complementary slackness conditions.
If $\frac{\mu} { M } - \zeta \nu_k\sum _{l \in \mathcal K} g_l (i)=0$, it is optimal to arbitrarily choose $q_k(i)$ in $[v_k(i), P_{\max}m_k(i)]$.
According to the above analysis, the optimal $q_k^* (i)$ can be expressed as}}
{\color{myc2}{
\begin{equation}\label{appentheorem1eem3eq2_3}
q_k^* (i) = \left\{
\begin{array}{ll}
 \! \! v_k (i) & \text{if} \; \frac{\mu} { M } > \zeta \nu_k\sum _{l \in \mathcal K} g_l (i)  \\
\!\! P_{\max} m_k(i)  & \text{if} \; \frac{\mu} { M } \leq  \zeta \nu_k\sum _{l \in \mathcal K} g_l (i)
\end{array}
\right.
\end{equation}
}}
for all $i\in\mathcal M, k\in\mathcal K$.

Based on (\ref{appentheorem1kkt1optv}) and the complementary slackness conditions, we can calculate the optimal $v_k^* (i)$ as
\begin{equation}\label{appentheorem1eem3eq5}
v_k ^* (i) \!=\! \left. \!\frac{   m_k (i) }{ (\ln 2) \zeta \nu_k   g_{k } (i) } \!-\! \frac{  \Gamma \sigma^2 m_k (i) } { g_k (i) } \right|^ {q_k (i)} _0\!
, \quad \forall i\in\mathcal M, k \in \mathcal K,
\end{equation}
where $a|_b^c=\min\{\max\{a,b\},c\}$.
Due to the fact that the objective function (\ref{eem3max1}a) is an increasing function of $v_k^*$, the optimal $q_k^* (i)$ and $v_k^* (i)$are given by (\ref{eem3eq2_3}) and (\ref{eem3eq5}), respectively.

\section{Proof of Theorem 3}
\setcounter{equation}{0}
\renewcommand{\theequation}{\thesection.\arabic{equation}}

Given ($ \pmb q, \bar { \pmb q }, \pmb v$), the Lagrange function of problem (\ref{eem3max0}) can be written by
\begin{eqnarray}
&&\!\!\!\!\!\!\!\!\!\mathcal L_3
(\pmb{m}, \pmb n,  \pmb \phi,   \pmb  \eta_0, \pmb \eta_1, \pmb \eta_2)
= \sum_{k\in \mathcal K}
U_{\alpha} ( \bar R_k^{\text {DL } } )
+ \sum_{k\in \mathcal K} U_{\alpha}  (\bar R_k^{\text{UL}})
\nonumber\\&&\!\!\!\!\!\!\!\!\!
+ {\color{myc2}{\sum_{i\in\mathcal M}\sum_{k \in \mathcal K} \eta_{k0} (i) \left( P_{\max}m_k(i)-  q_k(i)
\right)}}
\nonumber\\&&\!\!\!\!\!\!\!\!\! + \sum_{i\in\mathcal M} \phi (i) \left( 1- \sum_{k\in \mathcal K} ( m_k(i) + n_k (i) )
\right)
\nonumber\\&&\!\!\!\!\!\!\!\!\!+  \sum_{ i\in \mathcal M }   \sum _{k\in \mathcal K} \eta_{1k} (i) m _k (i)
+  \sum_{ i\in \mathcal M }  \sum _{k\in \mathcal K}  \eta_{2k} (i) n _k (i),
\end{eqnarray}
where $\bar R_k^{\text {DL}}$ and $\bar R_k ^{\text{UL}}$ are defined in (\ref{appenDeq1}) and ((\ref{appenDeq1_2})).
According to \cite{boyd2004convex}, the optimal solution of problem (\ref{eem3max0}) with given ($\pmb q, \bar { \pmb q }, \pmb v$) should satisfy:
\begin{subequations}\label{appenDkkt1}
\begin{align}
&\frac { \partial \mathcal L_3}
{\partial m_k(i)}= (\bar R_k ^{\text {DL}})^{-\alpha}
h\left(\frac{g_k (i) v_k (i)} { \Gamma \sigma^2 m_k (i) } \right)
 - \phi (i) +{\color{myc2}{\eta_{0k}(i)}}
  \nonumber\\
 &\qquad\qquad+\eta_{1k}(i)= 0 \\
& \frac { \partial \mathcal L_3}
{\partial n_k(i)}
=(\bar R_k ^{\text {UL}})^{-\alpha}
h \left(\frac{g_k (i) \bar q_k (i)}{\Gamma\sigma^2 n_k(i)}\right)
 - \phi (i)
 + \eta_{2k}(i) = 0,
\end{align}
\end{subequations}
for all $i \in \mathcal M$, $ k \in \mathcal K $.
By solving equations (\ref{appenDkkt1}) and using complementary slackness conditions, the optimal $m_k (i) $ and $ n_k (i) $ can be calculated as (\ref{Optalphamaxeq2}) and (\ref{Optalphamaxeq2_2}), respectively.
Substituting (\ref{Optalphamaxeq2}) and (\ref{Optalphamaxeq2_2}) into (\ref{eem3max0}b) with equality from Lemma 1, we find that $\phi (i)$ should satisfy (\ref{Optalphamaxeq2_3}).

With fixed ($ \pmb m, \pmb n$), the Lagrange function of problem (\ref{eem3max0}) can be written by
\begin{eqnarray}
&&\!\!\!\!\!\!\!\!\!\mathcal L_4
( \pmb q, \bar {\pmb q}, \pmb v, \pmb \nu, \mu, \pmb \eta_0, \pmb \eta_3, \pmb \eta_4, \pmb \eta_5  )
=
\nonumber\\&&\!\!\!\!\!\!\!\!\! \!  \sum_{k\in \mathcal K} U_{\alpha}  (\bar R_k^{ \text {DL} }) +
\sum_{k\in \mathcal K} U_{\alpha}  (\bar R_k^{\text{UL}})
\nonumber\\&&\!\!\!\!\!\!\!\!\! \! + \! \sum _{ i \in \mathcal M } \! \sum_{k \in \mathcal K}\! \nu_k \! \left( \! \zeta g_k(i) \sum_{l\in\mathcal K } q_l (i) -  \zeta g_k (i) v_k (i)
\right.
\nonumber\\&&\!\!\!\!\!\!\!\!\! \!
\left.
\!+\!
   {\color{myc2}{\zeta_0}} \sum_{l=1} ^{k-1} g_{lk} (i) \bar q_l (i)
\!+\!
   {\color{myc2}{\zeta_0}} \sum_{l=k+1}^K g_{lk} (i-1)  \bar q_l (i-1) \! - \! \bar q_k (i)
    \!  \right)
 \nonumber\\&&\!\!\!\!\!\!\!\!\!
    +  {\color{myc2}{\sum_{i\in\mathcal M}\sum_{k \in \mathcal K} \eta_{0k} (i) \left( P_{\max} m_k (i) - q_k(i)
\right)}}
\nonumber\\&&\!\!\!\!\!\!\!\!\! \!
  +
\mu\left( P_{\text {avg} } - \frac 1 M \sum_{i\in\mathcal M } \sum_{k \in \mathcal K} q_k(i)  \right)
 \nonumber\\&&\!\!\!\!\!\!\!\!\!
+
\sum_{ i\in \mathcal M } \sum _{k\in \mathcal K}  \eta_{4k} (i) (q _k (i) -v_k (i))
 + \sum_{ i\in \mathcal M } \sum _{k\in \mathcal K} \eta_{3k} (i) \bar q_k (i)
 \nonumber\\&&\!\!\!\!\!\!\!\!\! \!
+ \sum_{ i\in \mathcal M }   \sum _{k\in \mathcal K} \eta _{5k} (i)  v_k (i).
\end{eqnarray}
The optimal solution of problem (\ref{eem3max0}) with given ($ \pmb m, \pmb n$) should satisfy:
\begin{subequations}\label{appenDkkt2}
\begin{align}
 &\frac { \partial \mathcal  L_4}
{\partial q_k(i)}= - {\color{myc2}{\eta_{0k} (i)}} - \frac{\mu} { M } +\zeta \nu_k\sum _{l \in \mathcal K} g_l (i) + \eta_{ 4 k } (i) = 0 \\
& \frac { \partial \mathcal L_4 }
{ \partial \bar q_k (i) }
= \frac{ g_k (i) n_k (i) } { (\ln 2) ( \bar R_k ^{\text{UL}} ) ^{ \alpha } ( \Gamma \sigma^2 n_k (i) + g_k (i) \bar q_k (i) ) }
\nonumber \\ & \qquad \qquad
+ {\color{myc2}{\zeta_0}} \nu_k \sum_{ l \in \mathcal K \setminus \{k\}  } g_{kl} (i) - \nu_k + \eta_{ 3 k } (i) = 0 \\
& \frac{ \partial \mathcal L _4 }
{ v_k (i) }
= \frac{ g_k (i) m_k (i) } { (\ln 2) (\bar R_k ^{\text {UL}} )^{\alpha} ( \Gamma \sigma^2 m_k (i) + g_k (i) v_k (i) ) }
 \nonumber \\ & \qquad \qquad- \zeta \nu_k   g_{k } (i) +  \eta_{ 5k } (i) - \eta _{ 4k } (i) = 0,
\end{align}
\end{subequations}
for all $i \in \mathcal M$, $ k \in \mathcal K $.
Similar to (\ref{appentheorem1kkt5_1}) in Appendix C, the optimal $q_k (i) $ can be presented in (\ref{Optalphamaxeq3}).
According to (\ref{appenDkkt2}) and complementary slackness conditions, the optimal  $ \bar  q_k (i) $ and $ v_k (i)$ can be calculated as (\ref{Optalphamaxeq3_2}) and (\ref{Optalphamaxeq1}), respectively.


\section{Proof of Theorem 4}
\setcounter{equation}{0}
\renewcommand{\theequation}{\thesection.\arabic{equation}}

Given ($\pmb q, \bar { \pmb q }, \pmb v$), the Lagrange function of problem (\ref{fto3max2}) can be written by
\begin{eqnarray}
&&\!\!\!\!\!\!\!\!\! \mathcal L_5
(\theta, \pmb{m}, \pmb n, \pmb \psi_1, \pmb \psi_2, \pmb \phi, \pmb \eta_0, \pmb \eta_1, \pmb \eta_2)
=
\nonumber\\&&\!\!\!\!\!\!\!\!\!
\theta + \sum_{k \in \mathcal K} \psi_{1k} (
  \bar R_k^{\text {DL } } -\theta )
+ \sum_{k\in \mathcal K}   \psi_{ 2k } ( \bar R_k^{\text{UL}} -\theta )
\nonumber\\&&\!\!\!\!\!\!\!\!\!
+  \sum_{i\in\mathcal M} \phi (i) \! \left(\!  1\! -\!  \sum_{k\in \mathcal K}\!  ( m_k(i) + n_k (i) )
 \right)
 \nonumber\\&&\!\!\!\!\!\!\!\!\! +   {\color{myc2}{\sum_{i\in\mathcal M}\sum_{k \in \mathcal K}  \! \eta_{0k} (i)\!  \left(  P_{\max} m_k(i) - q_k(i)
\right)}}
\nonumber\\&&\!\!\!\!\!\!\!\!\!  +  \sum_{ i\in \mathcal M }   \sum _{k\in \mathcal K} \eta_{1k} (i) m _k (i)
 +  \sum_{ i\in \mathcal M }  \sum _{k\in \mathcal K}  \eta_{2k} (i) n _k (i),
\end{eqnarray}
where $ \pmb \psi_1= \{\psi_{1k}\}_{k \in \mathcal K} $  and   $ \pmb \psi _2 = \{\psi_{2k}\}_{k \in \mathcal K} $  are nonnegative Lagrangian multipliers associated with (\ref{fto3max2}b) and (\ref{fto3max2}c), respectively.
According to \cite{boyd2004convex}, the optimal solution of problem (\ref{fto3max2}) with fixed ($\pmb q, \bar { \pmb q }, \pmb v$) should satisfy:
\begin{subequations}\label{appenEkkt1}
\begin{align}
&\frac { \partial \mathcal L_5}
{ \partial \theta } = 1- \sum _{ k \in \mathcal K } ( \psi_{1k} + \psi _{2k } ) =0
\\
&\frac { \partial \mathcal L_5}
{\partial m_k(i)}= \psi_{1k}
h\left(\frac{g_k (i) v_k (i)} { \Gamma \sigma^2 m_k (i) } \right)
 - \phi (i) +{\color{myc2}{\eta_{0k}(i)}} \nonumber\\
 &\qquad\qquad +\eta_{1k}(i)= 0 \\
& \frac { \partial \mathcal L_5}
{\partial n_k(i)}
= \psi_{ 2k }
h \left(\frac{g_k (i) \bar q_k (i)}{\Gamma\sigma^2 n_k(i)}\right)
 - \phi (i) + \eta_{2k}(i) = 0,
\end{align}
\end{subequations}
for all $i \in \mathcal M$, $ k \in \mathcal K $.
By solving equations (\ref{appenEkkt1}b)-(\ref{appenEkkt1}c) and using complementary slackness conditions, the optimal $m_k (i) $ and $ n_k (i) $ can be calculated as (\ref{fto3eq2}) and (\ref{fto3eq2_2}), respectively.
Substituting (\ref{fto3eq2}) and (\ref{fto3eq2_2}) into (\ref{eem3max0}b) with equality from Lemma 1, we find that $\phi (i)$ should satisfy (\ref{fto3eq2_3}).

Given ($ \pmb m, \pmb n$), the Lagrange function of problem (\ref{eem3max0}) can be written by
\begin{eqnarray}
&&\!\!\!\!\!\!\!\!\!\!\!\mathcal L_6
( \theta, \pmb q, \bar {\pmb q}, \pmb v, \pmb \psi_1, \pmb \psi_2, \pmb \nu, \mu, \pmb \eta_0, \pmb \eta_3, \pmb \eta_4, \pmb \eta_5  ) =
\nonumber\\&&\!\!\!\!\!\!\!\!\!\theta + \sum_{k \in \mathcal K} \psi_{1k} (
  \bar R_k^{\text {DL } } -\theta )
+ \sum_{k\in \mathcal K}   \psi_{ 2k } ( \bar R_k^{\text{UL}} -\theta )
\nonumber\\&&\!\!\!\!\!\!\!\!\!\!\!
 \! + \! \sum_{k \in \mathcal K} \! \sum _{ i \in \mathcal M } \! \nu_k \! \left( \! \zeta g_k(i) \sum_{l\in\mathcal K } q_l (i) -  \zeta g_k (i) v_k (i)
 \right.\nonumber\\ &&\!\!\!\!\!\!\!\!\!\!\!
 \left.
\!+
   {\color{myc2}{\zeta_0}} \sum_{l=1} ^{k-1} g_{lk} (i) \bar q_l (i)
\!+
   {\color{myc2}{\zeta_0}} \sum_{l=k+1}^K g_{lk} (i-1)  \bar q_l (i-1) \! - \! \bar q_k (i)
    \!  \right) \nonumber\\&&\!\!\!\!\!\!\!\!\!\!\!
    +
\mu \left(  P_{\text {avg} } -  \frac 1 M \sum_{i\in\mathcal M }\!  \sum_{k \in \mathcal K}  q_k(i)   \right)
 \nonumber\\&&\!\!\!\!\!\!\!\!\!\!\!+   {\color{myc2}{\sum_{i\in\mathcal M}\sum_{k \in \mathcal K}  \! \eta_{0k} (i)\!  \left(  P_{\max} m_k(i) - q_k(i)
\right)}}  +
 \sum_{ i\in \mathcal M } \sum _{k\in \mathcal K} \eta_{3k} (i) \bar q_k (i)
\nonumber\\&&\!\!\!\!\!\!\!\!\!\!\!\!\!
\!+\!\sum_{ i\in \mathcal M } \sum _{k\in \mathcal K}  \eta_{4k} (i) (q _k (i) \!-\!v_k (i))\!+ \!\sum_{ i\in \mathcal M }   \sum _{k\in \mathcal K} \eta _{5k} (i)  v_k (i).
\end{eqnarray}
The optimal solution of problem (\ref{fto3max2}) with fixed ($ \pmb m, \pmb n$) should satisfy:
\begin{subequations}\label{appenEkkt2}
\begin{align}
&\frac { \partial \mathcal L_6}
{ \partial \theta } = 1- \sum _{ k \in \mathcal K } ( \psi_{1k} + \psi _{2k } ) =0 \\
  &\frac { \partial \mathcal  L_6 }
{\partial q_k(i)}= - {\color{myc2}{\eta_{0k}(i)}} - \frac{\mu} { M } +\zeta \nu_k\sum _{l \in \mathcal K} g_l (i) + \eta_{ 4 k } (i) = 0\\
& \frac { \partial \mathcal L_6 }
{ \partial \bar q_k (i) }
= \frac{ g_k (i)  \psi _{2k } n_k (i) } { (\ln 2)   ( \Gamma \sigma^2 n_k (i) + g_k (i) \bar q_k (i) ) } \nonumber \\ & \qquad\qquad + {\color{myc2}{\zeta_0}} \nu_k \sum_{ l \in \mathcal K \setminus \{k\}  } g_{kl} (i) - \nu_k + \eta_{ 3 k } (i) = 0 \\
& \frac{ \partial \mathcal L _6 }
{ v_k (i) }
= \frac{ g_k (i)  \psi_{1k} m_k (i) } { (\ln 2)  ( \Gamma \sigma^2 m_k (i) + g_k (i) v_k (i) ) }
\nonumber \\ & \qquad\qquad- \zeta \nu_k   g_{k } (i) +  \eta_{ 5k } (i) - \eta _{ 4k } (i) = 0,
\end{align}
\end{subequations}
for all $i \in \mathcal M$, $ k \in \mathcal K $.
Similar to (\ref{appentheorem1kkt5_1}) in Appendix C, the optimal $q_k (i) $ is given in (\ref{fto3eq3}).
According to (\ref{appenEkkt2}) and complementary slackness conditions, the optimal  $ \bar  q_k (i) $ and $ v_k (i)$ can be calculated as (\ref{fto3eq3_2}) and (\ref{fto3eq1}), respectively.

\bibliographystyle{IEEEtran}
\bibliography{IEEEabrv,MMM}

\end{document}